\documentclass[12pt]{article}
\usepackage{jheppub}

\pdfoutput=1

\usepackage{amsmath,bbm,array,amsfonts,graphicx,wrapfig,lscape,float,mathtools,multirow,longtable}
\usepackage[dvipsnames]{xcolor}
\usepackage{array}

\newcommand{\be}{\begin{equation}}
\newcommand{\ee}{\end{equation}}
\newcommand{\beq}{\begin{equation}}
\newcommand{\beql}[1]{\begin{equation}\label{#1}}
\newcommand{\eeq}{\end{equation}}
\newcommand{\ba}{\begin{array}}
\newcommand{\ea}{\end{array}}
\newcommand{\bea}{\begin{eqnarray}}
\newcommand{\beal}[1]{\begin{eqnarray}\label{#1}}
\newcommand{\eea}{\end{eqnarray}}
\newcommand{\ben}{\begin{enumerate}}
\newcommand{\een}{\end{enumerate}}
\newcommand{\bean}{\begin{eqnarray*}}
\newcommand{\eean}{\end{eqnarray*}}
\newcommand{\eref}[1]{(\ref{#1})}
\newcommand{\sref}[1]{\S\ref{#1}}
\newcommand{\tref}[1]{Table~\ref{#1}}
\newcommand{\nn}{\nonumber}

\newcommand{\fref}[1]{Figure \ref{#1}}
\newcommand{\btab}[1]{\begin{tabular}{#1}}
\newcommand{\etab}{\end{tabular}}

\def \Black {\pdfsetcolor{0 0 0 1}}

\newcommand{\comment}[1]{}

\newcommand{\ud}{\mathrm{d}}

\newcommand{\qed}{\nobreak \ifvmode \relax \else
      \ifdim\lastskip<1.5em \hskip-\lastskip
      \hskip1.5em plus0em minus0.5em \fi \nobreak
      \vrule height0.75em width0.5em depth0.25em\fi}

\definecolor{darkspringgreen}{rgb}{0.09, 0.45, 0.27}
\definecolor{forestgreen}{rgb}{0.13, 0.55, 0.13}

\usepackage{array}
\usepackage{physics}
\usepackage{subcaption}
\usepackage{tikz,tikz-3dplot}
\usepackage[colorlinks=true]{hyperref}
\newcolumntype{C}[1]{>{\centering\let\newline\\\arraybackslash\hspace{0pt}}m{#1}}

\definecolor{yellow2}{rgb}{0.98, 0.80, 0.20}

\title{On the Master Space for Brane Brick Models} 
\author[a]{Minsung Kho} 
\author[a,b]{and Rak-Kyeong Seong}

\affiliation[a]{
Department of Mathematical Sciences, and 
${}^{b}$Department of Physics,\\ 
Ulsan National Institute of Science and Technology,\\
50 UNIST-gil, Ulsan 44919, South Korea
}

\emailAdd{minsung@unist.ac.kr}	
\emailAdd{seong@unist.ac.kr}

\preprint{
\begin{flushright}
UNIST-MTH-23-RS-03 \\
\end{flushright}
}

\abstract{
We systematically study the master space of brane brick models that represent a large class of $2d$ $(0,2)$ quiver gauge theories.
These $2d$ $(0,2)$ theories are worldvolume theories of D1-branes that probe singular toric Calabi-Yau 4-folds. 
The master space is the freely generated space of chiral fields subject to the $J$- and $E$-terms and the non-abelian part of the gauge symmetry.
We investigate several properties of the master space for abelian brane brick models with $U(1)$ gauge groups.
For example, we calculate the Hilbert series, which allows us by using the plethystic programme to identify the generators and defining relations of the master space. 
By studying several explicit examples, we also show that the Hilbert series of the master space can be expressed in terms of characters of irreducible representations of the full global symmetry
of the master space.
}

\begin{document}

\maketitle

%=================================================================
\section{Introduction}

In this work, we take a peek at the rich algebro-geometric structure of the master space of $2d$ $(0,2)$ supersymmetric gauge theories that arise as worldvolume theories of D1-branes probing toric Calabi-Yau 4-folds.
These theories are realized by a Type IIA brane configuration that is connected to the D1-brane at the Calabi-Yau singularity via T-duality \cite{Franco:2015tna,Franco:2015tya}.
These brane configurations and the corresponding $2d$ $(0,2)$ gauge theories are known as brane brick models and led to new interesting developments in various contexts \cite{Franco:2016fxm,Franco:2016nwv,Franco:2016qxh,Franco:2021ixh,Franco:2018qsc}.

In particular, these recent developments involve the study of the mesonic moduli space of brane brick models \cite{Franco:2015tna,Franco:2015tya}.
The mesonic moduli space is defined as the space of gauge invariant operators under the $J$- and $E$-terms of the brane brick model. 
When the $2d$ theory has only $U(1)$ gauge groups, the mesonic moduli space is exactly the toric Calabi-Yau 4-fold associated to the brane brick model.
The gauge invariant operators carry charges under the mesonic flavor and $U(1)_R$ symmetries which combined become the isometry of the toric Calabi-Yau 4-fold.
Brane brick models and their mesonic moduli spaces have been recently fully classified for large classes of toric Calabi-Yau 4-folds such as brane brick models corresponding to smooth Fano 3-folds \cite{Franco:2022gvl} and cones over the Sasaki-Einstein 7-manifolds $Y^{p.k} (\mathbb{CP}^1 \times \mathbb{CP}^1)$ and $Y^{p.k} (\mathbb{CP}^2)$ \cite{Franco:2022isw}.

As introduced in \cite{Franco:2015tna}, in analogy to brane tilings \cite{Hanany:2005ve, Franco:2005rj, Hanany:2010zz, Forcella:2008bb, Forcella:2008eh, Forcella:2009bv, Zaffaroni:2008zz,Forcella:2008ng} realizing $4d$ $\mathcal{N}=1$ supersymmetric gauge theories associated to toric Calabi-Yau 3-folds, brane brick models have a master space that is larger than the mesonic moduli space.
The master space of the brane brick model is defined as the space of chiral fields subject to the $J$- and $E$-terms constraints and the non-abelian part of the gauge symmetry of the brane brick model, while the mesonic moduli space requires gauge invariance under the full gauge symmetry. 

In the case when the brane brick model has only $U(1)$ gauge groups, the master space simply takes the form of the following algebraic variety,
\beal{es01a01}
\mathcal{F}^\flat = 
\text{Spec} ~\mathbb{C}^{n_\chi} [X_1, \dots, X_{n_\chi}] / \langle J_a , E_a \rangle ~,~
\eea
where $n_\chi$ are the number of chiral fields $X_1, \dots, X_{n_\chi}$, and $J_a = 0$ and $E_a = 0$ are the $J$- and $E$-terms corresponding pairwise to a Fermi field $\Lambda_a$ of the brane brick model.
We note that the master space is a toric variety \cite{fulton,cox1995homogeneous,sturmfels1996grobner} given that the $J$- and $E$-terms are all binomial relations. 
Furthermore, we note that the variety in \eref{es01a01} is generally reducible into irreducible components. 
We call the top-dimensional irreducible component as the coherent component ${}^{\text{Irr}}\mathcal{F}^\flat$.
In the following work, when we refer to the master space of the brane brick model, we automatically refer to its irreducible coherent component ${}^{\text{Irr}}\mathcal{F}^\flat$. 

The $U(1)$ symmetries that are not imposed on the master space become symmetries along the additional directions that are added to the mesonic moduli space in order to form the master space. 
We refer to these added directions as baryonic directions and the
$U(1)$ symmetries as the baryonic part of the global symmetry of the master space for brane brick models.\footnote{These terms are borrowed from the master spaces of brane tilings \cite{Hanany:2005ve, Franco:2005rj, Hanany:2010zz, Forcella:2008bb, Forcella:2008eh, Forcella:2009bv, Zaffaroni:2008zz,Forcella:2008ng}.}
Out of the $G$ $U(1)$ symmetries in an abelian brane brick model, 
$G-1$ are independent because of the bifundamental nature of the chiral fields.
We have also the 3 global mesonic flavor symmetries and the $U(1)_R$-symmetry of the mesonic moduli space, which together with the 
independent $U(1)^{G-1}$ symmetries gives us the total rank of the global symmetry of the master space, $G+3$.
This also represents the dimension of the master space for abelian brane brick models with $U(1)$ gauge groups.

%--------------------------------------------------------
\begin{table}[ht!]
\begin{center}
\resizebox{1\hsize}{!}{
\begin{tabular}{|c|c|c|c|}
\hline
$\mathcal{M}^{mes}$ & ${}^{\text{Irr}}\mathcal{F}^\flat$ & $d(\mathcal{F}^\flat)$ & global symmetry of $\mathcal{F}^\flat$\\
\hline \hline
$\text{SPP}\times \mathbb{C}$ & 
$\mathcal{C} \times \mathbb{C}^3$ & 
6 & 
$SU(3) \times SU(2) \times SU(2) \times U(1) \times U(1)_R$
\\
\hline
$\mathbb{C}^4/\mathbb{Z}_4 ~(1,1,1,1)$ &
$[16,36]$ &
7 &
$SU(4) \times SU(4) \times U(1)_R$
\\
\hline
$Q_{1,1,1}$ &
$[10,6]$ &
7 &
$SU(4) \times SU(2) \times SU(2) \times U(1)_R$
\\
\hline
$Y^{2,4}(\mathbb{CP}^2)$ &
$[18,24]$ &
9 &
$SU(3) \times SU(2) \times SU(2) \times U(1)\times U(1)\times U(1)\times U(1)\times U(1)_R$
\\
\hline
\end{tabular}
}
\caption{
The mesonic moduli space $\mathcal{M}^{mes}$ of the brane brick model, the coherent component of the master space ${}^{\text{Irr}}\mathcal{F}^\flat$, the dimension of ${}^{\text{Irr}}\mathcal{F}^\flat$ and the full global symmetry of ${}^{\text{Irr}}\mathcal{F}^\flat$. 
The notation $[n_{g}, n_{r}]$ denotes the number of generators $n_g$ and the number of first order relations amongst the generators $n_r$ for ${}^{\text{Irr}}\mathcal{F}^\flat$ if it is a non-complete intersection. 
\label{texamples}
}
\end{center}
\end{table}
%--------------------------------------------------------

In the following work, we systematically calculate the Hilbert series \cite{Benvenuti:2006qr,Feng:2007ur,Butti:2006au,Butti:2007jv} for the master spaces of a selection of brane brick models corresponding to different toric Calabi-Yau 4-folds. 
The Hilbert series for the master space is a generating function that counts operators invariant under only the non-abelian part of the gauge symmetry and are subject to the $J$- and $E$-terms of the brane brick model.
In the following work, we will concentrate on master spaces of abelian brane brick models whose gauge groups are all $U(1)$.
Even with this restriction, we see that the master spaces exhibit extremely rich algebro-geometric properties.
For example, by the use of the plethystic programme \cite{hanany2007counting,Benvenuti:2006qr,Feng:2007ur,Butti:2006au,Butti:2007jv}, we obtain expressions for the generators and defining first order relations amongst the generators of the master space. 
By identifying the full enhanced global symmetry of the master space, we show how the generators and defining relations of the master spaces transform under the full global symmetry.
We also express the generators and relations of the master spaces in terms of GLSM fields that we obtain through the forward algorithm \cite{Feng:2000mi,Gulotta:2008ef,Franco:2015tna} for brane brick models.
\tref{texamples} summarizes the full global symmetry of the master spaces that we study in this work. 

By studying closely the geometric structure of the master spaces for brane brick models using the Hilbert series, we discover new phenomena specific to master spaces for brane brick models. 
These new discoveries involve the occurrence of extra GLSM fields that over-parameterize the master space for certain brane brick models, the enhancement of global symmetries of the master spaces, and the discovery that master spaces for brane brick models are toric but not necessarily Calabi-Yau. 
In the following work, we summarize these discoveries and illustrate their connection to the master spaces for brane brick models that correspond to $\text{SPP}\times \mathbb{C}$ \cite{Franco:2016fxm}, $\mathbb{C}^4/\mathbb{Z}_4$ with orbifold action $(1,1,1,1)$ \cite{Davey:2010px, Hanany:2010ne,Franco:2015tna}, $Q^{1,1,1}$ \cite{DAuria:1983sda, 1986PhR...130....1D,Nilsson:1984bj, Franco:2015tna, Franco:2015tya} and $Y^{2,4} (\mathbb{CP}^2)$ \cite{Martelli:2008rt, Gauntlett:2004hh,Franco:2022isw}.

Our work is organized as follows. In section \sref{sec:021}, we give a brief introduction to brane brick models and discuss in section \sref{sec:022} the forward algorithm \cite{Feng:2000mi,Gulotta:2008ef,Franco:2015tna} that allows us to construct the mesonic moduli spaces for brane brick models. 
We introduce the master space for brane brick models in section \sref{sec:031} and summarize the computation for the Hilbert series \cite{Benvenuti:2006qr,Feng:2007ur,Butti:2006au,Butti:2007jv} in section \sref{sec:032}, using either directly the $J$- and $E$-terms of the brane brick model or
the symplectic quotient description of the master space in terms of GLSM fields. 
Section \sref{sec:04} illustrates our findings in four explicit examples of master spaces for brane brick models. 
We conclude our work with a summary of our findings and a discussion for future research directions in section \sref{sec:05}.
\\

%=================================================================
\section{Background \label{sec:02}}

%--------------------------------------------------------
\subsection{Brane Brick Models \label{sec:021}}

The worldvolume theories of D1-branes probing a toric Calabi-Yau 4-fold form a large class of $2d$ $(0,2)$ supersymmetric gauge theories, which can be realized by a Type IIA brane configuration known as a brane brick model \cite{Franco:2015tna,Franco:2015tya}.
The brane brick model is connected to the D1-branes at the toric Calabi-Yau singularity by T-duality.
The Type IIA brane configuration consists of D4-branes wrapping a 3-torus $T^3$.
They are also suspended from a NS5-brane that wraps a holomorphic surface $\Sigma$ defined as,
\beal{es02a01}
\Sigma ~:~ P(x,y,z) = 0 ~,~
\eea
where $P(x,y,z)$ is the Newton polynomial in $x,y,z\in \mathbb{C}^*$ of the toric diagram associated to the probed toric Calabi-Yau 4-fold. 
The D4-brane meets the NS5-brane precisely where the holomorphic surface $\Sigma$ intersects the 3-torus $T^3$.
\tref{tbraneconfig} shows the Type IIA brane configuration for brane brick models realizing $2d$ $(0,2)$ supersymmetric gauge theories corresponding to toric Calabi-Yau 4-folds. 

%----------
\begin{table}[ht!]
\begin{center}
\begin{tabular}{|c|cc|cccccc|cc|}
\hline
\; & 0 & 1 & 2 & 3 & 4 & 5 & 6 & 7 & 8 & 9 \\
\hline \hline
D4 & $\times$ & $\times$ & $\cdot$ & $\times$ & $\cdot$ & $\times$ & $\cdot$ & $\times$ & $\cdot$ & $\cdot$\\
NS5 & $\times$ & $\times$ & \multicolumn{6}{c|}{------------$\Sigma$------------} & $\cdot$ & $\cdot$ \\
\hline
\end{tabular}
\caption{
The Type IIA brane configuration corresponding to a brane brick model on $T^3$.
\label{tbraneconfig}
}
\end{center}
\end{table}
%----------

The intersections between D4-brane and the NS5-brane form a tessellation of the 3-torus $T^3$.
This tessellation of the 3-torus is precisely what we call as the brane brick model.
The following summarizes the dictionary between components of the brane brick model on $T^3$ and the corresponding $2d$ $(0,2)$ supersymmetric gauge theory \cite{Franco:2015tya}: 
\begin{itemize}
\item \textbf{Bricks.}
These are 3-dimensional polytopes that form the fundamental building blocks of the tessellation of the 3-torus. Each brane brick corresponds to a $U(N)_i$ gauge group of the $2d$ supersymmetric gauge theory. 

\item \textbf{Faces.}
The boundary of a brane brick consists of even-sided 2-dimensional polygons.
A subset of these polygonal faces are oriented along their boundary edges. Such oriented faces correspond to bifundamental chiral fields $X_{ij}$ in the corresponding $2d$ $(0,2)$ theory. 
All other polygonal faces that are unoriented along their boundary edges correspond to Fermi fields $\Lambda_{ij}$ (and their conjugate $\bar{\Lambda}_{ij}$). 
These Fermi faces are always $4$-sided in a brane brick model. 
The two brane bricks adjacent to a polygonal face correspond to the two gauge groups $U(N)_i$ and $U(N)_j$ under which $X_{ij}$ or $\Lambda_{ij}$ associated to the brick face is in the bifundamental representation.

%--------------------------------
\begin{figure}[H]
\begin{center}
\resizebox{0.75\hsize}{!}{
\includegraphics[height=6cm]{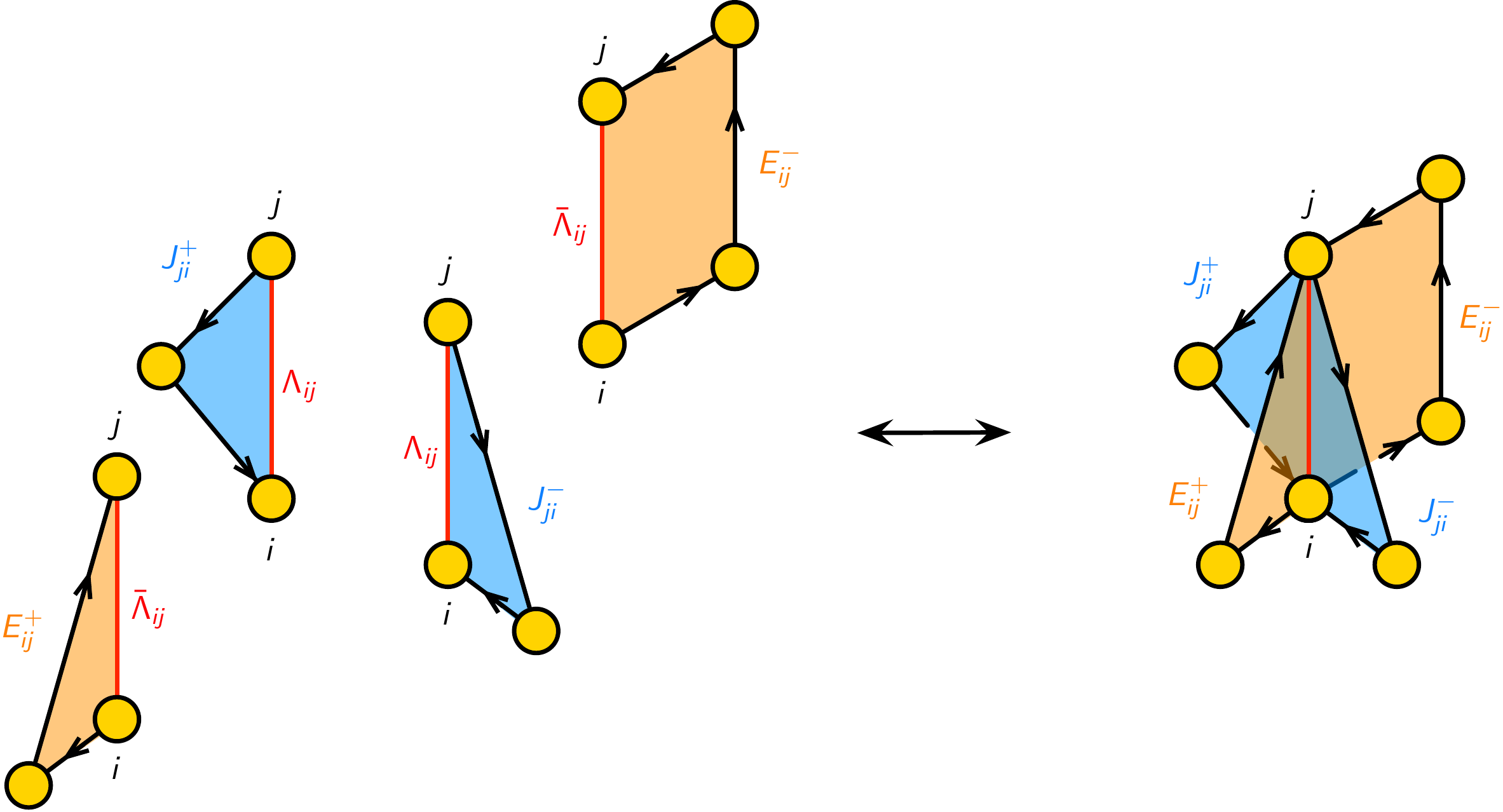} 
}
\caption{
The four plaquettes corresponding to $\Lambda_{ij}$ (and its conjugate $\bar{\Lambda}_{ij}$).
\label{fplaquettes}}
\end{center}
\end{figure}
%--------------------------------

\item \textbf{Edges.}
In a brane brick model, edges are always connected to a single brick face corresponding to $\Lambda_{ij}$ and a collection of oriented brick faces corresponding to a set of chiral fields $X_{ij}$. 
Accordingly, each brick edge is associated to monomial terms known as \textbf{plaquettes} in the brane brick model, which take one of the following forms
\beal{es02a10}
\Lambda_{ij} \cdot J^{+}_{ji} ~,~
\Lambda_{ij} \cdot J^{-}_{ji} ~,~
\bar{\Lambda}_{ij} \cdot E_{ij}^{+} ~,~
\bar{\Lambda}_{ij} \cdot E_{ij}^{-} ~,~
\eea
where $J^{\pm}_{ji}$ and $E_{ij}^{\pm}$ are monomial products of chiral fields. 
Plaquettes corresponding to opposite edges of the 4-sided Fermi face are identified to either a binomial $J$-term or a binomial $E$-term of the $2d$ $(0,2)$ supersymmetric gauge theory,
\beal{es02a11}
\ba{rccc}
\Lambda_{ij} : &J_{ji} &= & J_{ji}^{+} - J_{ji}^{-} ~,~\\
\bar{\Lambda}_{ij} : & E_{ij} &=& E_{ij}^{+} - E_{ij}^{-} ~,~
\ea
\eea
as illustrated in \fref{fplaquettes}.

\end{itemize}

In terms of the $J$- and $E$-terms of a brane brick model and the plaquettes associated to them, 
we are able to define special collections of chiral fields known as \textbf{brick matchings} \cite{Franco:2015tna,Franco:2015tya}.
A brick matching $p_\mu$ is a collection of chiral fields such that the chiral fields in the brick matching cover the plaquettes $(\Lambda_{ij} \cdot J_{ji}^{+},~ \Lambda_{ij} \cdot J_{ji}^{-})$ or the plaquettes $(\bar{\Lambda}_{ij} \cdot E_{ij}^{+},~ \bar{\Lambda}_{ij} \cdot E_{ij}^{-})$ exactly once each. 
We can summarize the chiral fields $X_m$ contained in a brick matching $p_\mu$ in terms of a brick matching matrix $P$, whose components take the following form,
\beal{es02a20}
P_{m\mu} = 
\left\{
\ba{ll}
1 & ~~~X_m \in p_\mu \\
0 & ~~~X_m \notin p_\mu \\
\ea
\right.
~.~
\eea
Additionally, the chiral fields $X_m$ of a brane brick model can be expressed in terms of products of brick matchings as follows, 
\beal{es02a21}
X_m = \prod_{\mu} p_{\mu}^{P_{m\mu}} ~.~
\eea
We note that brick matchings $p_\mu$ correspond to GLSM fields \cite{Witten:1993yc,Franco:2015tna,Franco:2015tya} that describe the mesonic moduli spaces of brane brick models. 

GLSM fields play an important role in the symplectic quotient description \cite{fulton,cox1995homogeneous} of the mesonic moduli space $\mathcal{M}^{mes}$ as well as the master space $\mathcal{F}^\flat$ of brane brick models. The following sections are going to give a brief overview of GLSM fields and their role in describing master spaces of brane brick models.
\\

%--------------------------------------------------------
\subsection{The Forward Algorithm \label{sec:022}}

The \textbf{forward algorithm} first introduced in \cite{Franco:2015tna} for brane brick models allows us to construct the mesonic moduli spaces of the corresponding $2d$ $(0,2)$ theories using GLSM fields. 
In the following section, we review the forward algorithm and illustrate the role played by GLSM fields to describe mesonic moduli spaces and master spaces of brane brick models. 
\\

\paragraph{GLSM Fields.}
A key step in the forward algorithm is the construction of the $P$-matrix in \eref{es02a20} encoding the GLSM fields $p_\mu$ in terms of chiral fields $X_m$ in the brane brick model.
We first observe that due to the binomial $J$- and $E$-terms of the $2d$ $(0,2)$ theory, 
\beal{es03a01}
(\Lambda_{ij}, \bar{\Lambda}_{ij}) ~:~ J_{ji} =  J_{ji}^{+} - J_{ji}^{-} = 0 ~,~ E_{ij} = E_{ij}^{+} - E_{ij}^{-} = 0 ~,~
\eea
where $J^{\pm}_{ji}$ and $E_{ij}^{\pm}$ are monomial products of chiral fields $X_m$, 
not all chiral fields in the brane brick model are independent to each other. 
In fact, by relabelling the independent fields as $v_k$, we can express all chiral fields $X_m$ as products of the following form, 
\beal{es03a02}
X_m = \prod_{k} v_k^{K_{mk}} ~,~
\eea
where $m=1,\dots, n_{\chi}$ labels the $n_\chi$ chiral fields $X_m$ in the brane brick model, and $k=1,\dots G+3$ labels the independent fields $v_k$ with $G$ being the number of gauge groups. 
Here, $K$ is a $n_\chi \times (G+3)$-dimensional matrix which identifies the binomial $J$- and $E$-terms with the independent fields $v_k$. 

Because the $J$- and $E$-terms of the brane brick model are binomial, they form a binomial ideal which is related to toric geometry \cite{sturmfels1996grobner}. 
In terms of toric geometry, the $K$-matrix in \eref{es03a02} defines a cone $\textbf{M}^+$, which is generated by non-negative linear combinations of the vectors $\vec{K}_m \in \mathbb{Z}^{G+3}$ encoded in the $K$-matrix.
The cone $\textbf{M}^+$ has a dual cone $\textbf{N}^+$ which is generated by non-negative linear combinations of another set of $(G+3)$-dimensional vectors $\vec{T}_\mu$ where $\mu=1,\dots,c$. 
These vectors can be combined to form a $(G+3)\times c$-dimensional matrix known as the $T$-matrix. 
We note that the cones $\textbf{M}^+$ and $\textbf{N}^+$ are dual to each other such that 
\beal{es03a05}
\vec{K} \cdot \vec{T} \geq 0 ~,~
\eea
which determines the number $c$ of distinct vectors $\vec{T}_\mu$. 

We can use now the $T$-matrix to identify the independent fields $v_k$ with a set of new fields $p_\mu$ 
such that the chiral fields can be expressed in terms of products of $p_\mu$ with strictly positive powers, 
\beal{es03a06}
v_ k = \prod_\mu p_{\mu}^{T_{k\mu}} ~.~
\eea
By combining the above expression with \eref{es03a02}, we obtain an expression of the $P$-matrix originally defined in \eref{es02a21} as follows, 
\beal{es03a07}
P_{n_\chi \times c} = K_{n_\chi \times (G+3)} \cdot T_{(G+3)\times c} ~,~
\eea
where the new fields $p_\mu$ with $\mu=1,\dots, c$ are GLSM fields of the brane brick model.
\\

\paragraph{The Mesonic Moduli Space and Toric Calabi-Yau 4-folds.}
Let us focus on abelian brane brick models with $U(1)$ gauge groups. 
The \textbf{mesonic moduli space} $\mathcal{M}^{mes}$ of the brane brick model \cite{Franco:2015tna,Franco:2015tya}
is the toric Calabi-Yau 4-fold geometry probed by a single D1-brane given that the worldvolume theory is a $2d$ $(0,2)$ supersymmetric gauge theory with $U(1)$ gauge groups. 
The geometry of the toric Calabi-Yau 4-fold is encoded in the brane brick model.
In order to obtain the geometry of the toric Calabi-Yau 4-fold from the brane brick model, we first express the $J$- and $E$-terms as well as the $D$-terms of the brane brick models as $U(1)$ charges on the GLSM fields:

\begin{itemize}
\item
\textbf{$J$- and $E$-terms.}
In terms of the GLSM fields summarized in the $P$-matrix, the $U(1)$ charges under the $J$- and $E$-terms of the brane brick model are given by the kernel of the $P$-matrix,
\beal{es03a11}
(Q_{JE})_{(c-G-3)\times c} = \ker P_{n_\chi \times c} ~,~
\eea
where the columns correspond to GLSM fields and the rows correspond to the $U(1)^{c-G-3}$ charges carried by the GLSM fields due to the $J$- and $E$-terms. Here $G$ refers to the number of gauge groups in the brane brick model.

\item
\textbf{$D$-terms.}
Similarly, the $D$-terms of the brane brick model 
are captured in terms of $U(1)$ charges carried by GLSM fields as follows,
\beal{es03a12}
\Delta_{(G-1)\times n_\chi} = (Q_{D})_{(G-1) \times c} \cdot P^t_{c\times n_\chi} ~,~
\eea
where $\Delta$ is the reduced incidence matrix of the quiver for the brane brick model. 
The columns of the $Q_D$-matrix correspond to the GLSM fields that carry the $U(1)^{G-1}$ charges from the $D$-terms of the brane brick model.

\end{itemize}

Overall, the GLSM fields carry in total $U(1)^{c-4}$ charges originating from both the $J$- and $E$-terms and the $D$-terms of the brane brick model. The total charge matrix is given by
\beal{es03a13}
(Q_{t})_{(c-4)\times c} = \left(
\ba{c}
Q_{JE}\\
Q_{D}
\ea
\right) ~.~
\eea
In terms of the total charge matrix $Q_t$, 
we can define the mesonic moduli space of an abelian brane brick model as the following symplectic quotient, 
\beal{es03a15}
\mathcal{M}^{mes} = \mathbb{C}^c // Q_t ~,~
\eea
where $\mathbb{C}^c$ is the freely generated space of GLSM fields, where $c$ is the number of GLSM fields in the brane brick model.
In terms of the total charge matrix $Q_t$, we can also define the coordinate matrix for the toric diagram of the toric Calabi-Yau 4-fold, 
\beal{es03a20}
(G_t)_{4\times c} = \ker (Q_t)_{(c-4)\times c} ~.~
\eea
The columns correspond to GLSM fields which map to vertices of the toric diagram.  
The coordinates of the vertices are given by 4-vertices in $\mathbb{Z}^4$.
Under the Calabi-Yau condition, the end-points of the 4-vertices are all on a 3-dimensional hyperplane in $\mathbb{Z}^4$, allowing us to draw the toric diagram as a convex lattice polytope on $\mathbb{Z}^3$. 
\\

\paragraph{Extra GLSM Fields.}
It was noted in \cite{Franco:2015tna}, that for certain brane brick models the toric diagram of the mesonic moduli space $\mathcal{M}^{mes}$ obtained from \eref{es03a20} exhibits additional vertices that are not on the 3-dimensional hyperplane like the rest of the vertices of the toric diagram.
That means, under a suitable $GL(4,\mathbb{Z})$ transformation on the coordinates $(x_1,x_2,x_3,x_4)\in \mathbb{Z}^4$ of the vertices of the toric diagram, the set of vertices splits into two. The first set contains vertices that are on a 3-dimensional hyperplane $GL(4,\mathbb{Z})$-transformed such that $x_4=1$, and all other vertices are outside the hyperplane with $x_4\neq 1$. 
These additional vertices with $x_4\neq 1$ correspond to what we call as \textbf{extra GLSM fields} \cite{Franco:2015tna}.

Although such extra GLSM fields manifest themselves as vertices in the toric diagram that seemingly break the Calabi-Yau condition on the mesonic moduli space $\mathcal{M}^{mes}$, it was shown in examples studied in previous work \cite{Franco:2015tna,Franco:2015tya} that the extra GLSM fields act as an over-parameterization of $\mathcal{M}^{mes}$.  
Given that the mesonic moduli space is parameterized by mesonic gauge invariant operators that can be expressed products of GLSM fields, the presence or absence of extra GLSM fields does not affect the spectrum of operators. 
This means that the generators and the first order defining relations formed by the generators of the quotient in \eref{es03a15} remain unaffected by the presence or absence of extra GLSM fields. 

In the following section, we describe the master space associated to a brane brick model.
Like the mesonic moduli space for brane brick models, the master space can be parameterized in terms of GLSM fields of the brane brick model. 
For certain brane brick models, the master spaces exhibit extra GLSM fields when the mesonic moduli space also contains extra GLSM fields. 
The extra GLSM fields in the mesonic moduli spaces however are not the same as the ones for the master space. 
For some examples, as we are going to see in section \sref{sec:04}, master spaces exhibit less extra GLSM fields than mesonic moduli spaces that contain extra GLSM fields. 
We are going to see in section \sref{sec:04} that extra GLSM fields of master spaces also have the features of an over-parameterization.
\\

%=================================================================
\section{Master Spaces for Brane Brick Models \label{sec:03}}

%--------------------------------------------------------
\subsection{An Introduction to the Master Space \label{sec:031}}

The \textbf{master space} $\mathcal{F}^\flat$ for brane brick models with $U(1)$ gauge groups takes the form of an algebraic variety \cite{Hanany:2010zz,Forcella:2008eh, Forcella:2009bv},
\beal{es04a01}
\mathcal{F}^\flat=
\text{Spec}~
\mathbb{C}^{n_\chi} [X_1, \dots, X_{n_\chi}] / \langle
J_a, E_a
\rangle
~,~
\eea
where $\mathcal{I}_{JE} =  \langle J_a, E_a \rangle$ is the quotienting ideal given by the relations $J_a = 0$ and $E_a = 0$ corresponding to all Fermi fields $\Lambda_a$.\footnote{Note that the $\text{Spec}$ of a coordinate ring $\mathbb{C} [x,y,z,w]$ gives the corresponding variety $\mathcal{X}$.
In a lot of the references as well as in this work, we interchangeably refer to the coordinate ring and its corresponding variety when we talk about mesonic moduli spaces and master spaces of supersymmetric gauge theories. }
This algebraic variety is analogous to the master space of $4d$ $\mathcal{N}=1$ supersymmetric gauge theories given by brane tilings \cite{Hanany:2005ve, Franco:2005rj, Hanany:2010zz, Forcella:2008bb, Forcella:2008eh, Forcella:2009bv, Zaffaroni:2008zz,Forcella:2008ng}, where the master space here is the space of vanishing $F$-terms. 

We can summarize the properties of the master space $\mathcal{F}^\flat$  for a brane brick model as follows:
\begin{itemize}

\item
Because the $J$- and $E$-terms are binomial relations in chiral fields of the brane brick model, the quotienting ideal $ \langle J_a, E_a \rangle$ is a binomial ideal and the resulting master space $\mathcal{F}^\flat$ is a \textbf{toric variety} \cite{fulton,cox1995homogeneous,sturmfels1996grobner}.

\item 
In general, the master space $\mathcal{F}^\flat$ is a reducible algebraic variety, which under \textbf{primary decomposition} \cite{M2,Gray:2008zs} can be decomposed into irreducible components. 
Amongst these irreducible components, there is a top-dimensional irreducible component which is of the same dimension and degree as $\mathcal{F}^\flat$.
We call this the \textbf{coherent component} of the master space and denote it by ${}^{\text{Irr}}\mathcal{F}^\flat$.
In the following discussion, whenever we refer to the master space of a brane brick model, we refer to the irreducible coherent component of the master space ${}^{\text{Irr}}\mathcal{F}^\flat$.

\item
The master space ${}^{\text{Irr}}\mathcal{F}^\flat$ can be expressed as a \textbf{symplectic quotient} of the following form,
\beal{es04a05}
{}^{\text{Irr}}\mathcal{F}^\flat
= 
\mathbb{C}^c // Q_{JE} ~,~
\eea
where $\mathbb{C}^c$ is parameterized by $p_1,\dots, p_c$, which are the GLSM fields of the brane brick model.
The $U(1)$ charges carried by the GLSM fields due to the $J$- and $E$-terms of the brane brick model are given by the $Q_{JE}$-matrix, which we defined in \eref{es03a11}. 

\item
The master space ${}^{\text{Irr}}\mathcal{F}^\flat$ is of \textbf{dimension} $G+3$, where $G$ refers to the number of $U(1)$ gauge groups in the abelian brane brick model.

\item 
For brane brick models with $U(N)$ gauge groups, the master space can be obtained by taking an additional quotient under the non-abelian $SU(N)$ part of the gauge symmetry, 
\beal{es04a05}
{}^{\text{Irr}}\mathcal{F}^\flat_N
= 
{}^{\text{Irr}}\mathcal{F}^\flat // SU(N)^G~,~
\eea
where $G$ is the number of $U(N)$ gauge groups in the brane brick model. 
The resulting \textbf{non-abelian master space} ${}^{\text{Irr}}\mathcal{F}^\flat_N$ is expected to be non-toric.
For the following work, we concentrate on master spaces ${}^{\text{Irr}}\mathcal{F}^\flat$ for abelian brane brick models with $U(1)$ gauge groups.

\end{itemize}

\paragraph{Global Symmetry of the Master Space.}
The master space ${}^{\text{Irr}}\mathcal{F}^\flat$ of an abelian brane brick model exhibits the following global symmetries:
\begin{itemize}
\item The \textbf{mesonic symmetry} of the master space is $U(1)^4$ or an enhancement with rank 4. 
The mesonic symmetry corresponds to the isometry of the mesonic moduli space, which is the toric Calabi-Yau 4-fold associated with the brane brick model. 
It contains the $U(1)_R$ symmetry and the mesonic flavor symmetries.

\item 
The $U(1)^G$ gauge symmetry of the brane brick model acts as a symmetry of the master space. 
This part of the global symmetry is known as the baryonic part of the global symmetry for the master space ${}^{\text{Irr}}\mathcal{F}^\flat$. 
We take the name from the master spaces for brane tilings and $4d$ $\mathcal{N}=1$ supersymmetric gauge theories \cite{Hanany:2010zz, Forcella:2008eh, Forcella:2009bv, Zaffaroni:2008zz,Forcella:2008ng}. 
In total, we have $G-1$ independent $U(1)$ symmetries because all chiral fields transform in the bifundamental or adjoint representation of the $U(1)$ in the quiver.
\end{itemize}

As a result, given the rank $G-1$ baryonic part of the global symmetry and the rank $4$ mesonic part of the global symmetry, the master space ${}^{\text{Irr}}\mathcal{F}^\flat$ as expected has a global symmetry of rank $G+3$. 
In terms of the brane brick model on the $3$-torus, we can identify the $U(1)^3$ part of the global symmetry of the master space ${}^{\text{Irr}}\mathcal{F}^\flat$ with the 3 $S^1$-cycles of the 3-torus of the brane brick model.
The remaining rank $G$ symmetry corresponds to the $G$ brane bricks of the brane brick model. 
\\

%--------------------------------------------------------
\subsection{The Hilbert Series of the Master Space \label{sec:032}}

\paragraph{Hilbert Series.}
A quintessential tool that is used to study the geometric structure of an algebraic variety is the \textbf{Hilbert series} \cite{Benvenuti:2006qr,Feng:2007ur,Butti:2006au,Butti:2007jv}.
Given an affine variety $Y$ in $\mathbb{C}^k$ over which $\mathcal{X}$ is a cone, we define the Hilbert series to be the generating function for the dimension of the graded pieces of the coordinate ring of the form
\beal{es06a01}
\mathbb{C}^k[x_1,\dots,x_k] / \langle f_i \rangle ~,~
\eea
where $\langle f_i \rangle$ is the quotienting ideal in terms of defining polynomials $f_i$ of $Y$.
The dimension of the $i$-th graded piece $Y_i$ is the number of algebraically independent degree $i$ polynomials on the variety $Y$.
Accordingly, the Hilbert series takes the form,
\beal{es06a02}
g(t; \mathcal{X}) = \sum_{i=0}^{\infty} \text{dim}_{\mathbb{C}} (Y_i)~ t^{i} ~,~
\eea
which always takes the form of a rational function. 
Here, the fugacity $t$ keeps track of the degree $i$ of the graded pieces $Y_i$.
In the case when the coordinate ring is multi-graded with pieces $Y_{\vec{i}}$ and grading $\vec{i}= (i_1,\dots,i_k)$, the Hilbert series takes a refined form as follows,
\beal{es06a03}
g(t_1,\dots,t_k; \mathcal{X}) = 
\sum_{\vec{i} = 0}^{\infty}
\text{dim}_{\mathbb{C}} (Y_{\vec{i}}) ~t_1^{i_1} \dots t_k^{i_k} ~,~
\eea
where the number of fugacities $t_i$ can be chosen to be as many as the dimension of the ambient space or as few as the dimension of $\mathcal{X}$ itself. 
\\

\paragraph{Hilbert Series of the Master Space.}
In our work, $\mathcal{X}$ is the master space $\mathcal{F}^\flat$ of a brane brick model with the corresponding coordinate ring given by,
\beal{es06a10}
\mathbb{C}^{n_\chi} [X_1, \dots, X_{n_\chi}] / \mathcal{I}_{JE}^{\text{Irr}}
~,~
\eea
where 
$X_1, \dots, X_{n_\chi}$ are the $n_\chi$ bifundamental chiral fields of the brane brick model.
$\mathcal{I}_{JE}^{\text{Irr}}$ is the ideal formed by the reduced $J$- and $E$-terms in the brane brick model which correspond to the coherent component of the master space ${}^{\text{Irr}}\mathcal{F}^\flat$.
The Hilbert series of ${}^{\text{Irr}}\mathcal{F}^\flat$
then can be obtained using the definition in \eref{es06a02}.
Given that the coordinate ring in \eref{es06a10} is in terms of chiral fields $X_1, \dots, X_{n_\chi}$ in the brane brick model, 
we can introduce a grading corresponding to the global symmetry charges on the chiral fields.
Taking $\bar{m}_1, \bar{m}_2$ as the fugacities for the mesonic flavor part of the global symmetry, $\bar{b}_1, \dots, \bar{b}_{G-1}$ as the baryonic part of the global symmetry, and $\bar{t}$ as the fugacity for the $U(1)_R$ symmetry,
a chiral field $X_a$ carrying charges under the mesonic flavor part $(q_1^a,q_2^a)$, charges under the baryonic part $(q_3^a, \dots, q_{G+2}^a)$, and a $U(1)_R$ charge $q_{G+3}^a$ can be associated with the following combination of fugacities,
\beal{es06a11}
y_a = \bar{m}_1^{q_1^a} \bar{m}_2^{q_2^a} ~ \bar{b}_1^{q_3^a} \cdots \bar{b}_{G-1}^{q_{G+2}^a} ~\bar{t}^{q_{G+3}}
~.~
\eea
Accordingly, the general form of the refined Hilbert series of the master space ${}^{\text{Irr}}\mathcal{F}^\flat$ based on \eref{es06a03} is as follows,
\beal{es06a12}
g(\bar{t}, \bar{m}_k,\bar{b}_l; {}^{\text{Irr}}\mathcal{F}^\flat) =
\sum_{\vec{i} = 0}^{\infty} n_{\vec{i}} ~ \bar{m}_1^{i_1} \bar{m}_2^{i_2} ~\bar{b}_1^{i_3} \cdots \bar{b}_{G-1}^{i_{G+2}}  ~ \bar{t}^{i_{G+3}} ~.~
\eea
The refined Hilbert series counts operators in terms of chiral fields that carry charges under the full global symmetry of the master space ${}^{\text{Irr}}\mathcal{F}^\flat$.
$n_{\vec{i}} \in \mathbb{Z}^{+}$ is the number of operators for a particular charge combination $\vec{i}$.
We can scale the fugacity $\bar{t}$ in such a way that it counts simply the overall degree of the operator.
Given the coordinate ring in \eref{es06a10} and its grading under the global symmetry of the master space ${}^{\text{Irr}}\mathcal{F}^\flat$, the corresponding refined Hilbert series in \eref{es06a12} can be obtained using \textit{Macaulay2} \cite{M2}. 

Alternatively, the Hilbert series of the master space ${}^{\text{Irr}}\mathcal{F}^\flat$ can be calculated using the symplectic quotient description of the master space in \eref{es04a05}.
Given the $Q_{JE}$-matrix from \eref{es03a11},
the refined Hilbert series of the master space ${}^{\text{Irr}}\mathcal{F}^\flat$ is defined by the Molien integral formula \cite{Butti:2007jv,Benvenuti:2006qr} as follows,
\beal{es06a20}
g(t_\alpha; {}^{\text{Irr}}\mathcal{F}^\flat)=
\prod_{i=1}^{|Q_{JE}|} \oint_{|z_i| =1}
\frac{\ud z_i}{2\pi i z_i} 
\prod_{\alpha=1}^{c} 
\frac{1}{1-t_\alpha \prod_{j=1}^{|Q_{JE}|} z_j^{(Q_{JE})_{j\alpha}}}
~,~
\eea
where $c$ is the number of GLSM fields $p_\alpha$ in the brane brick model and $|Q_{JE}|$ is the number of $U(1)$ charges encoded in the $Q_{JE}$-matrix. 
The fugacity $y_\alpha$ identifies the global symmetry charges carried by the GLSM field $p_\alpha$ as follows,
\beal{es06a21}
t_\alpha = m_1^{q^{\alpha}_{1}} m_2^{q^{\alpha}_2} ~b_1^{q^{\alpha}_{3}} \cdots b^{q^{\alpha}_{G+2}}_{G-1} ~ t
~,~
\eea
where the GLSM field $p_\alpha$ carries charges $(q^{\alpha}_1, q^{\alpha}_2)$ under the mesonic flavor part of the global symmetry and the charges $(q^{\alpha}_3, \dots, q^{\alpha}_{G+2})$ under the baryonic part of the global symmetry for the master space ${}^{\text{Irr}}\mathcal{F}^\flat$.
We can choose the fugacity $t$ to count the degree in GLSM fields $p_\alpha$ instead of the $U(1)_R$ charge. 

Note that the Hilbert series in \eref{es06a12} under fugacities for global symmetry charges carried by chiral fields $X_a$ is identical to the Hilbert series in \eref{es06a20} under the fugacities for global symmetry charges carried by GLSM fields since both Hilbert series describe the same master space ${}^{\text{Irr}}\mathcal{F}^\flat$.
The two Hilbert series can be mapped to each other under a fugacity map between fugacities in \eref{es06a21} and fugacities in \eref{es06a11} using the expression of chiral fields in \eref{es02a21} in terms of products of GLSM fields.
\\

\paragraph{Plethystics.}
The generators and the defining first order relations formed by the generators characterize the geometry of the master space ${}^{\text{Irr}}\mathcal{F}^\flat$. 
The \textbf{plethystic logarithm} \cite{hanany2007counting,Benvenuti:2006qr,Feng:2007ur,Butti:2006au,Butti:2007jv} of the Hilbert series of the master space ${}^{\text{Irr}}\mathcal{F}^\flat$ allows us to identify the generators and the first order relations formed by them. 
The plethystic logarithm is defined as 
\beal{es06a30}
\text{PL}[g(t_\alpha; {}^{\text{Irr}}\mathcal{F}^\flat) ]
=
\sum_{k=1}^{\infty}
\frac{\mu(k)}{k} \text{log}
~
\left[
g(t_\alpha^k;  {}^{\text{Irr}}\mathcal{F}^\flat)
\right]
~,~
\eea
where $\mu(k)$ is the M\"obius function, and $t_\alpha$ and $y_\alpha$ are fugacities of the Hilbert series corresponding to the GLSM fields $p_\alpha$ and the charges carried by them under the global symmetries of ${}^{\text{Irr}}\mathcal{F}^\flat$.

If the expansion of the plethystic logarithm is finite, the master space ${}^{\text{Irr}}\mathcal{F}^\flat$ is a \textbf{complete intersection} generated by a finite number of generators subject to a finite number of first order relations. 
In the case when the expansion is infinite, the master space ${}^{\text{Irr}}\mathcal{F}^\flat$ is a \textbf{non-complete intersection}, where the first order relations amongst the generators form higher order relations known as \textbf{syzygies} \cite{Benvenuti:2006qr,Feng:2007ur,Butti:2006au,Butti:2007jv}.
In the expansion of the plethystic logarithm, the first positive terms refer to the generators and the first negative terms in the expansion refer to the first order relations amongst the generators. 

%=================================================================
\section{Examples \label{sec:04}}

%=================================================================
\subsection{The Master Space for $\text{SPP} \times \mathbb{C}$ \label{sec:041}}
%=================================================================

\begin{figure}[H]
\begin{center}
\resizebox{0.3\hsize}{!}{
\includegraphics[height=6cm]{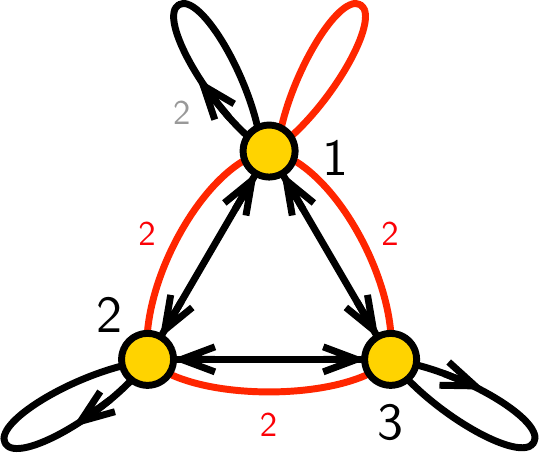} 
}
\caption{
The quiver diagram for the $\text{SPP} \times \mathbb{C}$ model. 
\label{f10a01}}
\end{center}
\end{figure}

The quiver diagram for the $\text{SPP} \times \mathbb{C}$ model \cite{Franco:2016fxm} is shown in \fref{f10a01}. The corresponding $J$- and $E$-terms take the following form,
\beal{es10a01}
\begin{array}{rrrclcrclcc}
& & & J  & & & & E  & & \\
\Lambda_{11} : & \ \ \ & X_{13} \cdot X_{31} &-& X_{12} \cdot X_{21} & \ \ \ \ &\Phi_{11} \cdot X_{11} &-& X_{11} \cdot \Phi_{11} \\
\Lambda_{21} : & \ \ \ & X_{12} \cdot X_{23} \cdot X_{32} &-& X_{11} \cdot X_{12}  & \ \ \ \ &\Phi_{22} \cdot X_{21} &-& X_{21} \cdot \Phi_{11} \\
\Lambda_{12} : & \ \ \ & X_{21} \cdot X_{11} &-& X_{23} \cdot X_{32} \cdot X_{21} & \ \ \ \ &\Phi_{11} \cdot X_{12} &-& X_{12} \cdot \Phi_{22} \\
\Lambda_{31} : & \ \ \ & X_{13} \cdot X_{32} \cdot X_{23}&-& X_{11} \cdot X_{13} & \ \ \ \ & \Phi_{33} \cdot X_{31} &-& X_{31} \cdot \Phi_{11} \\
\Lambda_{13} : & \ \ \ & X_{31} \cdot X_{11} &-& X_{32} \cdot X_{23} \cdot X_{31} & \ \ \ \ &\Phi_{11} \cdot X_{13} &-& X_{13} \cdot \Phi_{33} \\
\Lambda_{32} : & \ \ \ & X_{21} \cdot X_{12} \cdot X_{23} &-&X_{23} \cdot X_{31} \cdot X_{13}   & \ \ \ \ &\Phi_{33} \cdot X_{32} &-& X_{32} \cdot \Phi_{22}  \\
\Lambda_{23} : & \ \ \ & X_{32} \cdot X_{21} \cdot X_{12} &-& X_{31} \cdot X_{13} \cdot X_{32} & \ \ \ \ & \Phi_{22} \cdot X_{23} &-& X_{23} \cdot \Phi_{33} 
\end{array}
~,~
\eea
where we note that the $\text{SPP} \times \mathbb{C}$ model can be obtained by dimensional reduction of the $4d$ $\mathcal{N}=1$ supersymmetric gauge theory corresponding to the suspended pinch point (SPP) \cite{Morrison:1998cs,Franco:2005rj}.

%--------------------------------
%--------------------------------

\noindent
We can rewrite the $J$- and $E$-terms in terms of $G+3=6$ independent new variables, which are
\beal{es10a01a} 
v_1 = P_{11} ~,~
v_2 = X_{23} ~,~
v_3 = X_{32} ~,~
v_4 = X_{13} ~,~
v_5 = X_{21} ~,~
v_6 = X_{12} ~.~
\eea
These independent fields are related to the rest of the chiral fields in the $\text{SPP}\times \mathbb{C}$ model. This relationship is encoded in the following $K$-matrix,
\beal{es10a01b}
K= 
\resizebox{0.25\textwidth}{!}{$
\left(
\begin{array}{c|cccccc}
\; & v_1 & v_2 & v_3 & v_4 & v_5 & v_6
\\
\hline
\Phi_{11} & 1 & 0 & 0 & 0 & 0 & 0 \\
\Phi_{22} & 1 & 0 & 0 & 0 & 0 & 0 \\
\Phi_{33} & 1 & 0 & 0 & 0 & 0 & 0 \\
X_{11}     & 0 & 1 & 1 & 0 & 0 & 0 \\
X_{23}     & 0 & 1 & 0 & 0 & 0 & 0 \\
X_{32}     & 0 & 0 & 1 & 0 & 0 & 0 \\
X_{13}     & 0 & 0 & 0 & 1 & 0 & 0 \\
X_{31}     & 0 & 0 & 0 & -1 & 1 & 1 \\
X_{21}     & 0 & 0 & 0 & 0 & 1 & 0 \\
X_{12}     & 0 & 0 & 0 & 0 & 0 & 1 \\
\end{array}
\right)
$}
~.~
\eea
Using the forward algorithm, we obtain the $P$-matrix as follows,
\beal{es10a02}
P=
\resizebox{0.275\textwidth}{!}{$
\left(
\begin{array}{c|ccc|cc|cc}
\; & p_1 & p_2 & p_3 & p_4 & p_5 & p_6 & p_7 \\
\hline
\Phi_{11} & 0 & 0 & 1 & 0 & 0 & 0 & 0 \\
\Phi_{22} & 0 & 0 & 1 & 0 & 0 & 0 & 0 \\
\Phi_{33} & 0 & 0 & 1 & 0 & 0 & 0 & 0 \\
X_{11}     & 1 & 1 & 0 & 0 & 0 & 0 & 0 \\
X_{23}     & 1 & 0 & 0 & 0 & 0 & 0 & 0 \\
X_{32}     & 0 & 1 & 0 & 0 & 0 & 0 & 0 \\
X_{13}     & 0 & 0 & 0 & 1 & 0 & 1 & 0 \\
X_{31}     & 0 & 0 & 0 & 0 & 1 & 0 & 1 \\
X_{21}     & 0 & 0 & 0 & 1 & 0 & 0 & 1 \\
X_{12}     & 0 & 0 & 0 & 0 & 1 & 1 & 0 \\
\end{array}
\right)
$} ~,~
\eea
where $p_1,\dots,p_7$ are the GLSM fields of the brane brick model. 

In order to obtain the toric diagram for the master space ${}^{\text{Irr}}\mathcal{F}^\flat$
of the $\text{SPP}\times \mathbb{C}$ model, we first summarize the $U(1)$ charges on the
GLSM fields due to the $J$- and $E$-terms of the $\text{SPP}\times\mathbb{C}$ model. 
These charges are summarized in the $Q_{JE}$-matrix as follows,
\beal{es10a03}
Q_{JE}=
\resizebox{0.25\textwidth}{!}{$
\left(
\begin{array}{ccc|cc|cc}
p_1 & p_2 & p_3 & p_4 & p_5 & p_6 & p_7 \\
\hline
0 & 0 & 0 & -1 & -1 & 1 & 1 \\
\end{array}
\right)
$}
~,~
\eea
where we note that all GLSM fields carry charges under the $J$- and $E$-terms.
Additionally, the GLSM fields carry $U(1)$ charges due to the $D$-terms of the $\text{SPP}\times\mathbb{C}$ model. These charges are summarized in the $Q_D$-matrix as follows,
\beal{es10a03b}
Q_D=
\resizebox{0.25\textwidth}{!}{$
\left(
\begin{array}{ccc|cc|cc}
p_1 & p_2 & p_3 & p_4 & p_5 & p_6 & p_7 \\
\hline
 1 & -1 & 0 & 0 & -1 & 0 & 1 \\
 -1 & 1 & 0 & -1 & 0 & 0 & 1 \\
\end{array}
\right)
$}
~.~
\eea
We observe that the $Q_{JE}$- and $Q_{D}$-matrices together indicate that the mesonic moduli space $\mathcal{M}_{mes}$ has a completely broken symmetry of the form,
\beal{es10a04a}
U(1)_{f_1} \times U(1)_{f_2} \times U(1)_{f_3} \times U(1)_R
~.~
\eea
In comparison, when we focus on the master space ${}^{\text{Irr}}\mathcal{F}^\flat$, 
the $Q_{JE}$-matrix indicates that the global symmetry of ${}^{\text{Irr}}\mathcal{F}^\flat$ is enhanced to 
\beal{es10a04}
SU(3)_{(x_1,x_2)} \times SU(2)_y \times SU(2)_z \times U(1)_b \times U(1)_R ~,~
\eea
where the total rank of the global symmetry of the master space ${}^{\text{Irr}}\mathcal{F}^\flat$ is as expected $G+3=6$. 
Here we note that the $SU(3)$ enhancement in the global symmetry of ${}^{\text{Irr}}\mathcal{F}^\flat$ is due to the fact that the GLSM fields $(p_1,p_2,p_3)$ carry the same $Q_{JE}$ charges.
Furthermore, the $SU(2)$ enhancements in the global symmetry are due to the fact that the GLSM fields $(p_4,p_5)$ and $(p_6,p_7)$ carry the same $Q_{JE}$ charges. 
The charges on the GLSM fields due to the global symmetry in \eref{es10a04} are summarized in \tref{tex01}.

%--------------------------------
\begin{table}[ht!]
\centering
\begin{tabular}{|c|c|c|c|c|c|c|}
\hline
\; & $SU(3)_{(x_1,x_2)}$ & $SU(2)_{y}$ &$SU(2)_{z}$& $U(1)_{b}$ & $U(1)_{R}$ & fugacity\\
\hline\hline
$p_1$ & $(+1,~~0)$ & $0$ & $0$ &$+2$&$r_1$ &$t_1 = x_1 b^2 t$  \\
$p_2$ & $(-1,+1)$ & $0$ & $0$ &$+2$&$r_2$&$t_2 = x_1^{-1} x_2 b^2 t$  \\
$p_3$ & $(0,~-1)$ & $0$ & $0$ &$+2$&$r_3$&$t_3 = x_2^{-1} b^2 t $  \\
\hline
$p_4$ & $0$ & $+1$& $0$ & $-3$& $r_4$ &    $t_4 = y b^{-3} t$  \\
$p_5$ & $0$  & $-1$ & $0$ & $-3$ & $r_5$ &    $t_5= y^{-1} b^{-3} t$  \\
\hline
$p_6$ & $0$ & $0$  & $+1$&$0$ &$r_6$ &    $t_6 = z t$  \\
$p_7$ & $0$ & $0$  & $-1$ &$0$ & $r_7$ &    $t_7= z^{-1} t$  \\
\hline
\end{tabular}
\caption{The global symmetry charges of the master space ${}^{\text{Irr}}\mathcal{F}^\flat$ on the GLSM fields $p_\alpha$ for the $\text{SPP}\times\mathbb{C}$ model.
\label{tex01}}
\end{table}
%--------------------------------

Given that the $J$- and $E$-terms in \eref{es10a01} are all binomial, the master space ${}^{\text{Irr}}\mathcal{F}^\flat$ is toric and has a $5$-dimensional toric diagram which is given by
\beal{es10a05}
G_t^{{}^{\text{Irr}}\mathcal{F}^\flat}=
\resizebox{0.25\textwidth}{!}{$
\left(
\begin{array}{ccc|cc|cc}
\;  p_1 & p_2 & p_3 & p_4 & p_5 & p_6 & p_7 \\
\hline
 0 & 0 & 0 & 1 & 0 & 0 & 1 \\
 0 & 0 & 0 & 1 & 0 & 1 & 0 \\
 0 & 0 & 0 & -1 & 1 & 0 & 0 \\
 0 & 0 & 1 & 0 & 0 & 0 & 0 \\
 0 & 1 & 0 & 0 & 0 & 0 & 0 \\
 \hline
 1 & 1 & 1 & 1 & 1 & 1 & 1 \\
\end{array}
\right)
$}
~,~
\eea
where the GLSM fields $p_1,\dots,p_7$ correspond to extremal vertices of the toric diagram.

Using the symplectic quotient description of the master space ${}^{\text{Irr}}\mathcal{F}^\flat$
and the corresponding formula in \eref{es06a20} for the Hilbert series, 
we obtain the Hilbert series of ${}^{\text{Irr}}\mathcal{F}^\flat$ as follows,
\beal{es10a20}
g(
t_\alpha;
{}^{\text{Irr}}\mathcal{F}^\flat
)=\frac{1-t_4 t_5 t_6 t_7 }{\left(1-t_1\right) \left(1-t_2\right) \left(1-t_3\right) \left(1-t_4 t_6\right)
   \left(1-t_5 t_6\right) \left(1-t_4 t_7\right) \left(1-t_5 t_7\right)}
~,~
\nn\\
\eea
where the fugacities $t_\alpha$ correspond to the GLSM fields $p_\alpha$ in the brane brick model.
By taking all fugacities to be $t_\alpha = t$, we can write down the unrefined Hilbert series of the master space ${}^{\text{Irr}}\mathcal{F}^\flat$,
\beal{es10a21}
g(
t;
{}^{\text{Irr}}\mathcal{F}^\flat
)
=
\frac{1 + t^2}{
(1 - t)^3 (1 - t^2)^3
} ~,~
\eea
where the palindromic numerator indicates that the master space ${}^{\text{Irr}}\mathcal{F}^\flat$ is Calabi-Yau.
As a result, the master space ${}^{\text{Irr}}\mathcal{F}^\flat$ is a 6-dimensional toric Calabi-Yau manifold.
When we calculate the plethystic logarithm, 
\beal{es10a22}
\text{PL}\left[
g(
t_\alpha;
{}^{\text{Irr}}\mathcal{F}^\flat
)
\right]
= t_1 + t_2 + t_3 + t_4 t_6 + t_5 t_6 + t_4 t_7 + t_5 t_7 - t_4 t_5 t_6 t_7 ~,~
\eea
we further note that the master space ${}^{\text{Irr}}\mathcal{F}^\flat$ is a complete intersection.

Using the $P$-matrix, we can express the chiral fields of the brane brick model as products of GLSM fields as follows, 
\beal{es10a30}
&
\Phi_{11} = p_3 ~,~
\Phi_{22} = p_3 ~,~
\Phi_{33} = p_3 ~,~
X_{11} = p_1 p_2 ~,~
X_{23} = p_1 ~,~
X_{32} = p_2 ~,~
&
\nn\\
&
X_{13} = p_4 p_6 ~,~
X_{31} = p_5 p_7 ~,~
X_{21} = p_4 p_7 ~,~
X_{12} = p_5 p_6 ~.~
&
~.~
\eea
When we define a coordinate ring in terms of chiral fields, we can assign every chiral field a grading that corresponds to the degree of GLSM fields $p_\alpha$ in the expressions in \eref{es10a30}.
Under primary decomposition of the $J$- and $E$-terms in \eref{es10a01}, the coherent component takes the following form,
\beal{es10a35}
\mathcal{I}_{JE}^{\text{Irr}} =
\left\langle
\Phi_{22}-\Phi_{33}~,~
\Phi_{11}-\Phi_{33}~,~
X_{13} X_{31}-X_{21}X_{12}~,~ 
X_{23} X_{32}-X_{11}
\right\rangle
~.~
\eea
The master space ${}^{\text{Irr}}\mathcal{F}^\flat$ is then given by 
\beal{es10a36}
{}^{\text{Irr}}\mathcal{F}^\flat
= 
\text{Spec}~
\mathbb{C}^{10}[
\Phi_{ii},
X_{ij}
]
/ \mathcal{I}_{JE}^{\text{Irr}}
~.~
\eea
Using \textit{Macaulay2} \cite{M2}, with the chiral fields graded in terms of the corresponding degree in GLSM fields $p_\alpha$, we obtain as expected exactly the same Hilbert series as in \eref{es10a21}.

Using the following fugacity map,
\beal{es10a40}
&
t = t_1^{1/7} t_2^{1/7} t_3^{1/7} t_4^{1/7} t_5^{1/7} t_6^{1/7} t_7^{1/7} ~,~
x_1 = \frac{t_1^{2/3}}{{t_2}^{1/3}{t_3}^{1/3}}~,~ 
x_2 = \frac{{t_1^{1/3}}{t_2^{1/3}}}{t_3^{2/3}}~,~ 
&
\nn\\
&
y = \frac{t_4^{1/2}}{t_5^{1/2}}~,~
z = \frac{t_6^{1/2}}{t_7^{1/2}}~,~
b =\frac{t_1^{2/21} t_2^{2/21} t_3^{2/21}}{{t_4^{1/14}} {t_5^{1/14}} {t_6^{1/14}} {t_7^{1/14}}}~,~
&
\eea
we can rewrite the refined Hilbert series in \eref{es10a20} in terms of characters of irreducible representations of the global symmetry of the master space ${}^{\text{Irr}}\mathcal{F}^\flat$.
The refined Hilbert series in terms of characters of irreducible representations of the global symmetry takes the following form,
\beal{es10a41}
g (t,x_i,y,z,b;{}^{\text{Irr}}\mathcal{F}^\flat )
=
\sum_{n_1,n_2=0}^\infty [n_1,0;n_2;n_2] b^{2 n_1-3 n_2} t^{n_1+2n_2}
~,~
\eea
where $[m_1,m_2;n;k]=[m_1,m_2]_{SU(3)_{(x_1,x_2)}} [n]_{SU(2)_y}  [k]_{SU(2)_z}$.
The corresponding highest weight generating function \cite{Hanany:2014dia} is given by,
\beal{es10a42}
h\left(\mu_i,\nu,\kappa,b,t,{}^{\text{Irr}}\mathcal{F}^\flat\right)=\frac{1}{\left(1-\mu_1 ~b^2 t\right) \left(1-\nu \kappa ~b^{-3}t^2\right) }
\eea
where $ \mu_1^{m_1} \mu_2^{m_2} \nu^{n} \kappa^{k}  \sim [m_1,m_2]_{SU(3)} [n]_{SU(2)_{y}} [k]_{SU(2)_{z}}$.
The plethystic logarithm of the refined Hilbert series takes the form,
\beal{es10a45}
PL[g(t,x_i,y,z,b;{}^{\text{Irr}}\mathcal{F}^\flat)]= [1,0;0;0]  b^2 t +[0,0;1;1] b^{-3} t^2 -b^{-6} t^4
~.~
\eea
From the plethystic logarithm, we identify the generators of the master space ${}^{\text{Irr}}\mathcal{F}^\flat$ as,
\beal{es10a46}
A_i = (X_{23}~,~ X_{32}~,~ \Phi_{11}=\Phi_{22}=\Phi_{33})  &~~\leftrightarrow~~& 
+ [1,0;0;0]  b^2 t
~,~
\\
B_{jk} = \left(
\ba{cc}
X_{13} & X_{21} \\
X_{12} & X_{31}
\ea
\right)_{jk}
&~~\leftrightarrow~~&
+[0,0;1;1] b^{-3} t^2
~.~
\eea
The single relation at order $-b^{-6} t^4$ is given by,
\beal{es10a47}
\det B = 0 &~~\leftrightarrow~~& -b^{-6} t^4
~,~
\eea
where we note that the subspace $\mathbb{C}^4[B_{jk}]/\langle \det B = 0 \rangle$ corresponds to the conifold $\mathcal{C}$ \cite{Candelas:1989ug,Candelas:1989js}.
As a result, we identify the master space ${}^{\text{Irr}}\mathcal{F}^\flat$ of the $\text{SPP}\times \mathbb{C}$ brane brick model as the following 6-dimensional product space,
\beal{es10a50}
{}^{\text{Irr}}\mathcal{F}^\flat =  \mathcal{C} \times \mathbb{C}^3 
~,~
\eea
where $\mathbb{C}^3$ is generated by $A_i$ and the conifold $\mathcal{C}$ is generated by $B_{jk}$.
\tref{tex02} summarizes the generators of the master space ${}^{\text{Irr}}\mathcal{F}^\flat$ with the corresponding charges under the global symmetry of ${}^{\text{Irr}}\mathcal{F}^\flat$.
\\

%--------------------------------
\begin{table}[ht!]
\centering
\resizebox{0.95\textwidth}{!}{
\begin{tabular}{|c|c|c|c|c|c|c|c|}
\hline
generators  & GLSM fields & $SU(3)_{(x_1,x_2)}$ & $SU(2)_{y}$ & $SU(2)_{z}$ & $U(1)_{b}$ & $U(1)_R$\\
\hline\hline
$A_1=X_{23}$ & $p_1 $ & $\left(+1,~~0\right)$ &$0$&$0$& $+2$ & $r_1$\\
$A_2=X_{32}$ & $p_2 $ & $\left(-1,+1\right)$ &$0$&$0$&$+2$ & $r_2$\\
$A_3=\Phi_{11}=\Phi_{22}=\Phi_{33}$ & $p_3 $ & $\left(0,~+1\right)$ &$0$&$0$& $+2$ & $r_3$\\
\hline
$B_{11}=X_{13}$ & $p_4 p_6 $ & $\left(0,~~~0\right)$ &$+1$&$+1$&$-3$ & $r_4+r_6$\\
$B_{22}=X_{31}$ & $p_5 p_7 $ & $\left(0,~~~0\right)$ &$-1$&$-1$&$-3$ & $r_5+r_7$\\
$B_{12}=X_{21}$ & $p_4 p_7 $ & $\left(0,~~~0\right)$ &$+1$&$-1$&$-3$  & $r_4+r_7$\\
$B_{21}=X_{12}$ & $p_5 p_6 $ & $\left(0,~~~0\right)$ &$-1$&$+1$&$-3$  & $r_5+r_6$\\
\hline
\end{tabular}}
\caption{
Generators of the master space ${}^{\text{Irr}}\mathcal{F}^\flat$ for the $\text{SPP}\times\mathbb{C}$ model with the global symmetry charges. 
\label{tex02}
}
\end{table}
%--------------------------------

%======================================================================
\subsection{The Master Space for $\mathbb{C}^4/\mathbb{Z}_4 ~(1,1,1,1)$ \label{sec:042}}
%======================================================================

\begin{figure}[H]
\begin{center}
\resizebox{0.26\hsize}{!}{
\includegraphics[height=6cm]{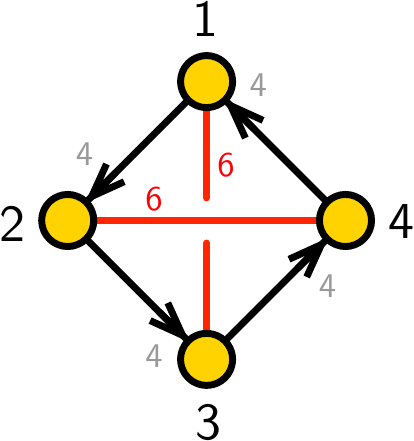} 
}
\caption{
The quiver diagram for the $\mathbb{C}^4/\mathbb{Z}_4 ~(1,1,1,1)$. 
\label{f11a01}}
\end{center}
\end{figure}

The $J$- and $E$-terms of the brane brick model for $\mathbb{C}^4/\mathbb{Z}_4 ~(1,1,1,1)$ \cite{Davey:2010px, Hanany:2010ne,Franco:2015tna} take the following form,
\beal{es11a01}
\begin{array}{rrclrcl}
 &  & J &  & & E &  \\
\Lambda_{13}^{1}  : & Z_{34} \cdot Y_{41}  &-& Y_{34} \cdot Z_{41} &   P_{12} X_{23}  &-&  X_{12} P_{23} \\ 
\Lambda_{13}^{2}  : & X_{34} \cdot Z_{41}  &-&  Z_{34}\cdot  X_{41}  &   P_{12} Y_{23}  &-&  Y_{12} P_{23} \\ 
\Lambda_{13}^{3}  : & Y_{34} \cdot X_{41}  &-& X_{34} \cdot Y_{41}  &   P_{12} Z_{23}  &-&  Z_{12} P_{23} \\ 
\Lambda_{24}^{1}  : & Z_{41} \cdot Y_{12}  &-& Y_{41} \cdot Z_{12}  &   P_{23} X_{34}  &-&  X_{23} P_{34} \\ 
\Lambda_{24}^{2}  : & X_{41} \cdot Z_{12}  &-&  Z_{41} \cdot X_{12}  &   P_{23} Y_{34}  &-&  Y_{23} P_{34} \\ 
\Lambda_{24}^{3}  : & Y_{41} \cdot X_{12}  &-& X_{41} \cdot Y_{12} &   P_{23} Z_{34}  &-&  Z_{23} P_{34} \\ 
\Lambda_{31}^{1}  : & Z_{12} \cdot Y_{23}  &-& Y_{12} \cdot Z_{23}  &   P_{34} X_{41}  &-&  X_{34} P_{41} \\ 
\Lambda_{31}^{2}  : & X_{12} \cdot Z_{23}  &-&  Z_{12} \cdot X_{23}  &   P_{34} Y_{41}  &-&  Y_{34} P_{41} \\ 
\Lambda_{31}^{3}  : & Y_{12} \cdot X_{23}   &-& X_{12} \cdot Y_{23} &   P_{34} Z_{41}  &-&  Z_{34} P_{41} \\ 
\Lambda_{42}^{1}  : & Z_{23} \cdot Y_{34}  &-& Y_{23}\cdot  Z_{34}  &   P_{41} X_{12}  &-&  X_{41} P_{12} \\ 
\Lambda_{42}^{2}  : & X_{23} \cdot Z_{34}  &-&  Z_{23} \cdot X_{34}  &   P_{41} Y_{12}  &-&  Y_{41} P_{12} \\ 
\Lambda_{42}^{3}  : &  Y_{23} \cdot X_{34} &-& X_{23} \cdot Y_{34}  &   P_{41} Z_{12}  &-&  Z_{41} P_{12}
 \end{array} 
~.~
\eea
The corresponding quiver diagram is shown in \fref{f11a01}.

Because the binomial $J$- and $E$-terms are not all independent, we can rewrite them in terms of 
$G+3=7$ independent new variables, which are
\beal{es11a01a} 
v_1 = P_{12} ~,~
v_2 = P_{23} ~,~
v_3 = X_{12} ~,~
v_4 = X_{34} ~,~
v_5 = X_{41} ~,~
v_6 = Y_{12} ~,~
v_7 = Z_{12} ~.~
\nn\\
\eea
These independent variables are then used to rewrite the $J$- and $E$-terms.
This is encoded in the $K$-matrix, which takes the following form
\beal{es11a01b}
K= 
\resizebox{0.25\textwidth}{!}{$
\left(
\begin{array}{c|ccccccc}
\; & v_1 & v_2 & v_3 & v_4 & v_5 & v_6 & v_7
\\
\hline
P_{12} &1 & 0 & 0 & 0 & 0 & 0 & 0 \\
P_{23} &0 & 1 & 0 & 0 & 0 & 0 & 0 \\
P_{34} &1 & 0 & -1 & 1 & 0 & 0 & 0 \\
P_{41} &1 & 0 & -1 & 0 & 1 & 0 & 0 \\
\hline
X_{12} &0 & 0 & 1 & 0 & 0 & 0 & 0 \\
X_{23} &-1 & 1 & 1 & 0 & 0 & 0 & 0 \\
X_{34} &0 & 0 & 0 & 1 & 0 & 0 & 0 \\
X_{41} &0 & 0 & 0 & 0 & 1 & 0 & 0 \\
\hline
Y_{12} &0 & 0 & 0 & 0 & 0 & 1 & 0 \\
Y_{23} &-1 & 1 & 0 & 0 & 0 & 1 & 0 \\
Y_{34} &0 & 0 & -1 & 1 & 0 & 1 & 0 \\
Y_{41} &0 & 0 & -1 & 0 & 1 & 1 & 0 \\
\hline
Z_{12} &0 & 0 & 0 & 0 & 0 & 0 & 1 \\
Z_{23} &-1 & 1 & 0 & 0 & 0 & 0 & 1 \\
Z_{34} &0 & 0 & -1 & 1 & 0 & 0 & 1 \\
Z_{41} &0 & 0 & -1 & 0 & 1 & 0 & 1 \\
\end{array}
\right)
$}
~.~
\eea
Following the forward algorithm, we obtain the $P$-matrix
\beal{es11a02}
 P=
 \resizebox{0.275\textwidth}{!}{$
\left(
\begin{array}{r|cccc|cccc}
\; & p_1 & p_2 & p_3 & p_4 & p_5 & p_6 & p_7 & p_8 \\
\hline
P_{12} & 1 & 0 & 0 & 0 & 1 & 0 & 0 & 0 \\
P_{23} & 1 & 0 & 0 & 0 & 0 & 1 & 0 & 0 \\
P_{34} & 1 & 0 & 0 & 0 & 0 & 0 & 1 & 0 \\
P_{41} & 1 & 0 & 0 & 0 & 0 & 0 & 0 & 1 \\
\hline
X_{12} & 0 & 1 & 0 & 0 & 1 & 0 & 0 & 0 \\
X_{23} & 0 & 1 & 0 & 0 & 0 & 1 & 0 & 0 \\
X_{34} & 0 & 1 & 0 & 0 & 0 & 0 & 1 & 0 \\
X_{41} & 0 & 1 & 0 & 0 & 0 & 0 & 0 & 1 \\
 \hline
Y_{12} & 0 & 0 & 1 & 0 & 1 & 0 & 0 & 0 \\
Y_{23} & 0 & 0 & 1 & 0 & 0 & 1 & 0 & 0 \\
Y_{34} & 0 & 0 & 1 & 0 & 0 & 0 & 1 & 0 \\
Y_{41} & 0 & 0 & 1 & 0 & 0 & 0 & 0 & 1 \\
 \hline
Z_{12} & 0 & 0 & 0 & 1 & 1 & 0 & 0 & 0 \\
Z_{23} & 0 & 0 & 0 & 1 & 0 & 1 & 0 & 0 \\
Z_{34} & 0 & 0 & 0 & 1 & 0 & 0 & 1 & 0 \\
Z_{41} & 0 & 0 & 0 & 1 & 0 & 0 & 0 & 1 \\
\end{array}
\right) 
$}
~,~
\eea
where $p_1,\dots,p_8$ are the GLSM fields of the $\mathbb{C}^4/\mathbb{Z}_4 ~(1,1,1,1)$ model. 

In order to obtain the toric diagram for the master space ${}^{\text{Irr}}\mathcal{F}^\flat$, we first identify the $U(1)$ charges on the GLSM fields due to the $J$- and $E$-terms of the $\mathbb{C}^4/\mathbb{Z}_4 ~(1,1,1,1)$ model.
These charges are summarized in the $Q_{JE}$-matrix as follows,
\beal{es11a03}
Q_{JE}=
\resizebox{0.265\textwidth}{!}{$
\left(
\begin{array}{cccc|cccc}
p_1 & p_2 & p_3 & p_4 & p_5 & p_6 & p_7 & p_8 \\
\hline
 1 & 1 & 1 & 1 & -1 & -1 & -1 & -1 \\
\end{array}
\right)
$}
~,~
\eea
where we note that all GLSM fields carry charges under the $J$- and $E$-terms.
The GLSM fields also carry $U(1)$ charges coming from the $D$-terms of the brane brick model, which are summarized in the $Q_D$-matrix as follows,
\beal{es11a03b}
Q_D=
\resizebox{0.265\textwidth}{!}{$
\left(
\begin{array}{cccc|cccc}
p_1 & p_2 & p_3 & p_4 & p_5 & p_6 & p_7 & p_8 \\
\hline
 0 & 0 & 0 & 0 & -1 & 1 & 0 & 0 \\
 0 & 0 & 0 & 0 &  0 &-1 & 1 & 0 \\
 0 & 0 & 0 & 0 &  0 & 0 &-1 & 1 \\
\end{array}
\right)
$}
~.~
\eea

%--------------------------------
%--------------------------------

The $Q_{JE}$- and $Q_{D}$-matrices together indicate that the mesonic moduli space $\mathcal{M}_{mes}$ has a symmetry of the form,
\beal{es11a04a}
SU(4)_{(x_1,x_2,x_3)}  \times U(1)_R
~.~
\eea
The $SU(4)$ enhancement is due to the GLSM fields $p_1,\dots, p_4$ carrying the same $Q_{JE}$- and $Q_{D}$-charges.
When we focus on the master space ${}^{\text{Irr}}\mathcal{F}^\flat$, 
the $Q_{JE}$-matrix indicates that the global symmetry of ${}^{\text{Irr}}\mathcal{F}^\flat$ is of the form
\beal{es11a04}
SU(4)_{(x_1,x_2,x_3)} \times SU(4)_{(y_1,y_2,y_3)} \times U(1)_R ~,~
\eea
where the total rank of the global symmetry of the master space ${}^{\text{Irr}}\mathcal{F}^\flat$ is as expected $G+3=7$. 
Here the two $SU(4)$ enhancements are due to the fact that the GLSM fields $(p_1,\dots, p_4)$ and $(p_5,\dots,p_8)$ carry the same $Q_{JE}$-charges.
We note that the $SU(4)$ mesonic flavor symmetry in \eref{es11a04a} is carried over to the global symmetry of the master space ${}^{\text{Irr}}\mathcal{F}^\flat$.
The second $SU(4)$ factor in \eref{es11a04} can be identified as the baryonic part of the global symmetry of the master space ${}^{\text{Irr}}\mathcal{F}^\flat$.
\tref{tex03} summarizes the charges on the GLSM fields due to the global symmetry of the master space ${}^{\text{Irr}}\mathcal{F}^\flat$.

%--------------------------------
\begin{table}[ht!]
\centering
\begin{tabular}{|c|c|c|c|c|}
\hline
\; & $SU(4)_{(x_1,x_2,x_3)}$ & $SU(4)_{(y_1,y_2,y_3)}$ & $U(1)_R$ & fugacity\\
\hline\hline
$p_1$ & $(+1,~0,~0)$ & $0$     & $r_1$ &   $t_1=x_1 t$  \\
$p_2$ & $(-1,+1,0) $ & $0$     & $r_2$ &   $t_2=x_1^{-1} x_2 t$  \\
$p_3$ & $(0,-1,+1) $ & $0$     & $r_3$ &   $t_3=x_2^{-1} x_3 t$  \\
$p_4$ & $(0,~~0,-1) $ & $0$     & $r_4$ &   $t_4=x_3^{-1} t$  \\
\hline
$p_5$ & $0$ & $(+1,~0,~0)$     & $r_5$ &   $t_5=y_1 t$  \\
$p_6$ & $0$ & $(-1,+1,0) $     & $r_6$ &   $t_6=y_1^{-1} y_2 t$  \\
$p_7$ & $0$ & $(0,-1,+1) $     & $r_7$ &   $t_7=y_2^{-1} y_3 t$  \\
$p_8$ & $0$ & $(0,~~0,-1) $     & $r_8$ &   $t_8=y_3^{-1} t$  \\
\hline
\end{tabular}
\caption{The global symmetry charges of the master space ${}^{\text{Irr}}\mathcal{F}^\flat$ on the GLSM fields $p_\alpha$ for the $\mathbb{C}^4/\mathbb{Z}_4 ~(1,1,1,1)$~[$\mathbb{P}^3$,~$\langle0\rangle$].
\label{tex03}}
\end{table}

Because the $J$- and $E$-terms in \eref{es11a01} are all binomial, the master space ${}^{\text{Irr}}\mathcal{F}^\flat$ is toric and has a $6$-dimensional toric diagram which is given by
\beal{es11a05}
G_t^{{}^{\text{Irr}}\mathcal{F}^\flat}=
\resizebox{0.265\textwidth}{!}{$
\left(
\begin{array}{cccc|cccc}
\;  p_1 & p_2 & p_3 & p_4 & p_5 & p_6 & p_7 & p_8 \\
\hline
 1 & 0 & 0 & 0 & 0 & 0 & 0 & 1 \\
 0 & 1 & 0 & 0 & 0 & 0 & 0 & 1 \\
 0 & 0 & 1 & 0 & 0 & 0 & 0 & 1 \\
 0 & 0 & 0 & 1 & 0 & 0 & 0 & 1 \\
 0 & 0 & 0 & 0 & 1 & 0 & 0 & -1 \\
 0 & 0 & 0 & 0 & 0 & 1 & 0 & -1 \\
 \hline
 1 & 1 & 1 & 1 & 1 & 1 & 1 &  1 \\
\end{array}
\right)
$}
~,~
\eea
where the GLSM fields $p_1,\dots,p_8$ correspond to extremal vertices of the toric diagram.

The symplectic quotient description of the master space ${}^{\text{Irr}}\mathcal{F}^\flat$
and the corresponding formula in \eref{es06a20} for the Hilbert series gives us the Hilbert series for ${}^{\text{Irr}}\mathcal{F}^\flat$ as follows,
\beal{es11a20}
&&
g(
t_\alpha;
{}^{\text{Irr}}\mathcal{F}^\flat
)
=
\frac{
P(t_\alpha)
}{
(1- t_1 t_5 )  (1-t_1 t_6) (1-t_1 t_7 ) (1-t_1 t_8) (1- t_2 t_5 )  (1- t_2 t_6) (1-t_2 t_7)
}\nn\\
&&
\hspace{1cm}
\times 
\frac{1}{
(1-t_2 t_8) (1-t_3 t_5 )  (1-t_3 t_6 ) (1-t_3 t_7 ) (1-t_3 t_8 ) (1-t_4 t_5)  (1-t_4 t_6) 
}
\nn\\
&&
\hspace{1cm}
\times 
\frac{1}{
(1-t_4 t_7)  (1-t_4 t_8)
}
~,~
\eea
where $P(t_\alpha)$ is the numerator of the Hilbert series and the fugacities $t_\alpha$ count the degrees in GLSM fields $p_\alpha$.

We can unrefine the Hilbert series by taking all fugacities to be $t_\alpha = t$.  
This gives us
\beal{es11a21}
&&
g(
t;
{}^{\text{Irr}}\mathcal{F}^\flat
)
=
\nn\\
&&
\hspace{1cm}
\frac{1-36 t^4+160 t^6-315 t^8+288 t^{10}-288 t^{14}+315 t^{16}-160 t^{18}+36 t^{20}-t^{24}}{(1-t^2)^{16}}
~,~
\nn\\
\eea
where the palindromic numerator indicates that the master space ${}^{\text{Irr}}\mathcal{F}^\flat$ is Calabi-Yau.
Accordingly, the master space ${}^{\text{Irr}}\mathcal{F}^\flat$ is a 7-dimensional toric Calabi-Yau manifold.
When we calculate the plethystic logarithm, 
\beal{es11a22}
&&
\text{PL}\left[
g(
t_\alpha;
{}^{\text{Irr}}\mathcal{F}^\flat
)
\right]
= 
(t_1 t_5+t_2 t_5+t_3 t_5+t_4 t_5+t_1 t_6+t_2 t_6+t_3 t_6+t_4 t_6+t_1 t_7+t_2 t_7
\nn\\
&&
\hspace{1cm}
+t_3 t_7+t_4 t_7+t_1 t_8+t_2 t_8+t_3 t_8+t_4 t_8)-(t_1 t_2 t_5 t_6+t_1 t_3 t_5 t_6+t_1 t_4 t_5 t_6
+t_2 t_3 t_5 t_6
\nn\\
&&
\hspace{1cm}
+t_2 t_4 t_5 t_6+t_3 t_4 t_5 t_6+t_1 t_2 t_5 t_7+t_1 t_3 t_5 t_7+t_1 t_4 t_5 t_7+t_2 t_3 t_5 t_7
+t_2 t_4 t_5 t_7+t_3 t_4 t_5 t_7
\nn\\
&&
\hspace{1cm}
+t_1 t_2 t_5 t_8+t_1 t_3 t_5 t_8+t_1 t_4 t_5 t_8+t_2 t_3 t_5 t_8+t_2 t_4 t_5 t_8
+t_3 t_4 t_5 t_8+t_1 t_2 t_6 t_7+t_1 t_3 t_6 t_7
\nn\\
&&
\hspace{1cm}
+t_1 t_4 t_6 t_7+t_2 t_3 t_6 t_7+t_2 t_4 t_6 t_7+t_3 t_4 t_6 t_7
+t_1 t_2 t_6 t_8+t_1 t_3 t_6 t_8+t_1 t_4 t_6 t_8+t_2 t_3 t_6 t_8
\nn\\
&&
\hspace{1cm}
+t_2 t_4 t_6 t_8+t_3 t_4 t_6 t_8+t_1 t_2 t_7 t_8
+t_1 t_3 t_7 t_8+t_1 t_4 t_7 t_8+t_2 t_3 t_7 t_8+t_2 t_4 t_7 t_8+t_3 t_4 t_7 t_8)
\nn\\
&&
\hspace{1cm}
+\dots~,~
\eea
we obtain an infinite series which indicates that the master space ${}^{\text{Irr}}\mathcal{F}^\flat$ is a non-complete intersection.

We can express the chiral fields of the brane brick model as products of GLSM fields using the $P$-matrix in \eref{es11a02}, 
\beal{es11a30}
&
P_{12} = p_1 p_5 ~,~
P_{23} = p_1 p_6 ~,~
P_{34} = p_1 p_7 ~,~
P_{41} = p_1 p_8 ~,~
&
\nn\\
&
X_{12} = p_2 p_5 ~,~
X_{23} = p_2 p_6 ~,~
X_{34} = p_2 p_7 ~,~
X_{41} = p_2 p_8 ~,~
&
\nn\\
&
Y_{12} = p_3 p_5 ~,~
Y_{23} = p_3 p_6 ~,~
Y_{34} = p_3 p_7 ~,~
Y_{41} = p_3 p_8 ~,~
&
\nn\\
&
Z_{12} = p_4 p_5 ~,~
Z_{23} = p_4 p_6 ~,~
Z_{34} = p_4 p_7 ~,~
Z_{41} = p_4 p_8 ~.~
&
\eea
For the coordinate ring in terms of chiral fields, we can assign every chiral field a grading that corresponds to the degree of GLSM fields $p_\alpha$ in the expressions in \eref{es11a30}.
The coherent component can be obtained using primary decomposition of the $J$- and $E$-terms in \eref{es11a01}.
It takes the following form,
\beal{es11a35}
&&
\mathcal{I}_{JE}^{\text{Irr}} =
\langle
~
X_{34} Y_{41}- Y_{34} X_{41}~,~ 
Y_{34} Z_{41}- Z_{34} Y_{41}~,~ 
Z_{34} X_{41}- X_{34} Z_{41}~,~ 
\nn\\
&&
\hspace{1.5cm}
X_{34} P_{41}- P_{34} X_{41}~,~ 
Y_{34} P_{41}- P_{34} Y_{41}~,~ 
Z_{34} P_{41}- P_{34} Z_{41}~,~
\nn\\
&&
\hspace{1.5cm}
X_{23} Y_{34}- Y_{23} X_{34}~,~ 
Y_{23} Z_{34}- Z_{23} Y_{34}~,~ 
Z_{23} X_{34}- X_{23} Z_{34}~,~ 
\nn\\
&&
\hspace{1.5cm}
X_{23} P_{34}- P_{23} X_{34}~,~ 
Y_{23} P_{34}- P_{23} Y_{34}~,~ 
Z_{23} P_{34}- P_{23} Z_{34}~,~ 
\nn\\
&&
\hspace{1.5cm}
X_{41} Y_{12}- Y_{41} X_{12}~,~ 
Y_{41} Z_{12}- Z_{41} Y_{12}~,~ 
Z_{41} X_{12}- X_{41} Z_{12}~,~ 
\nn\\
&&
\hspace{1.5cm}
X_{41} P_{12}- P_{41} X_{12}~,~ 
Y_{41} P_{12}- P_{41} Y_{12}~,~ 
Z_{41} P_{12}- P_{41} Z_{12}~,~ 
\nn\\
&&
\hspace{1.5cm}
X_{12} Y_{23}- Y_{12} X_{23} ~,~
Y_{12} Z_{23}- Z_{12} Y_{23}~,~ 
Z_{12} X_{23}- X_{12} Z_{23}~,~ 
\nn\\
&&
\hspace{1.5cm}
X_{12} P_{23}- P_{12} X_{23} ~,~
Y_{12} P_{23}- P_{12} Y_{23}~,~ 
Z_{12} P_{23}- P_{12} Z_{23}~,~ 
\nn\\
&&
\hspace{1.5cm}
X_{12} Y_{34}- Y_{12} X_{34}~,~ 
Y_{12} Z_{34}- Z_{12} Y_{34}~,~ 
Z_{12} X_{34}- X_{12} Z_{34} ~,~ 
\nn\\
&&
\hspace{1.5cm}
X_{12} P_{34}- P_{12} X_{34}~,~ 
Y_{12} P_{34}- P_{12} Y_{34}~,~ 
Z_{12} P_{34}- P_{12} Z_{34}~,~ 
\nn\\
&&
\hspace{1.5cm}
X_{41} Y_{23}- Y_{41} X_{23} ~,~ 
Y_{41} Z_{23}- Z_{41} Y_{23} ~,~  
Z_{41} X_{23}- X_{41} Z_{23} ~,~
\nn\\
&&
\hspace{1.5cm}
X_{41} P_{23}- P_{41} X_{23} ~,~
Y_{41} P_{23}- P_{41} Y_{23} ~,~ 
Z_{41} P_{23} -P_{41} Z_{23}
~
\rangle
~.~
\eea
The master space ${}^{\text{Irr}}\mathcal{F}^\flat$ is then given in terms of the ideal in \eref{es11a35} as follows,
\beal{es11a36}
{}^{\text{Irr}}\mathcal{F}^\flat
= 
\text{Spec}~
\mathbb{C}^{16}[
P_{ij},
X_{ij},
Y_{ij},
Z_{ij}
]
/ \mathcal{I}_{JE}^{\text{Irr}}
~,~
\eea
where the coordinate ring is in terms of the $16$ chiral fields of the brane brick model. 
By grading the chiral fields in terms of their corresponding degree in GLSM fields $p_\alpha$, we can use \textit{Macaulay2} \cite{M2} in order to obtain 
the Hilbert series for the master space ${}^{\text{Irr}}\mathcal{F}^\flat$.
As expected the Hilbert series takes exactly the same form as in \eref{es11a21}.

The following fugacity map,
\beal{es11a40}
&
t = t_1^{1/8} t_2^{1/8} t_3^{1/8} t_4^{1/8} t_5^{1/8} t_6^{1/8} t_7^{1/8} t_8^{1/8}~,~
x_1 = \frac{t_1^{3/4}}{t_2^{1/4} t_3^{1/4} t_4^{1/4}}~,~ 
x_2 = \frac{t_1^{1/2} t_2^{1/2}}{t_3^{1/2} t_4^{1/2}}~,~ 
x_3 = \frac{t_1^{1/4} t_2^{1/4} t_3^{1/4}}{t_4^{3/4}}~,~ 
&
\nn\\
&
y_1 = \frac{t_5^{3/4}}{t_6^{1/4} t_7^{1/4} t_8^{1/4}}~,~
y_2 = \frac{t_5^{1/2} t_6^{1/2}}{t_7^{1/2} t_8^{1/2}}~,~ 
y_3 = \frac{t_5^{1/4} t_6^{1/4} t_7^{1/4}}{t_8^{3/4}}~,~
&
\eea
allows us to rewrite the refined Hilbert series in \eref{es11a20} in terms of characters of irreducible representations of the global symmetry of the master space ${}^{\text{Irr}}\mathcal{F}^\flat$ in \eref{es11a04}.
The refined Hilbert series in terms of characters of irreducible representations of the global symmetry takes the following form,
\beal{es11a41}
g (t,x_i,y_j;{}^{\text{Irr}}\mathcal{F}^\flat )
=
\sum_{n=0}^\infty [n,0,0;n,0,0] t^{2n}
~,~
\eea
where $[m_1,m_2,m_3;n_1,n_2,n_3]=[m_1,m_2,m_3]_{SU(4)_{(x_1,x_2,x_3)}} [n_1,n_2,n_3]_{SU(4)_{(y_1,y_2,y_3)}}$.
The highest weight generating function \cite{Hanany:2014dia} takes the form,
\beal{es11a42}
h\left(t,\mu_i,\nu_j;{}^{\text{Irr}}\mathcal{F}^\flat\right)=\frac{1}{1-\mu_1 \nu_1 t^2 }
\eea
where $ \mu_1^{m_1} \mu_2^{m_2} \mu_3^{m_3} \nu_1^{n_1} \nu_2^{n_2} \nu_3^{n_3} \sim [m_1,m_2,m_3]_{SU(4)_{(x_1,x_2,x_3)}} [n_1,n_2,n_3]_{SU(4)_{(y_1,y_2,y_3)}}$.
In terms of characters of irreducible representations of the global symmetry, the plethystic logarithm of the refined Hilbert series of ${}^{\text{Irr}}\mathcal{F}^\flat$ takes the form,
\beal{es11a45}
PL[g(t,x_i,y_j;{}^{\text{Irr}}\mathcal{F}^\flat)]= [1,0,0;1,0,0] t^2 -[0,1,0;0,1,0] t^4 +\dots ~.~
\eea
From the plethystic logarithm, we identify the generators of the master space ${}^{\text{Irr}}\mathcal{F}^\flat$ as,
\beal{es11a46}
A_{ij} = \left(
\ba{cccc}
P_{12} & P_{23} & P_{34} & P_{41} \\
X_{12} & X_{23} & X_{34} & X_{41} \\
Y_{12} & Y_{23} & Y_{34} & Y_{41} \\
Z_{12} & Z_{23} & Z_{34} & Z_{41} 
\ea
\right)_{ij}
&~~\leftrightarrow~~&
+[1,0,0;1,0,0] t^2
~.~
\eea
The relation at order $-[0,1,0;0,1,0] t^4$ is given by,
\beal{es11a47}
N^{klmn}=\epsilon^{kli_1i_2} \epsilon^{mnj_1j_2}M_{i_1 j_1}M_{i_2 j_2}=0
 &~~\leftrightarrow~~& -[0,1,0;0,1,0] t^4
~,~
\eea
where $N^{klmn}=-N^{lkmn}=-N^{klnm}$.
\tref{tex04} summarizes the generators of the master space ${}^{\text{Irr}}\mathcal{F}^\flat$ with the corresponding global symmetry charges.
\\

%--------------------------------
\begin{table}[ht!]
\centering
\resizebox{0.75\textwidth}{!}{
\begin{tabular}{|c|c|c|c|c|c|}
\hline
generators  & GLSM fields & $SU(4)_{(x_1,x_2,x_3)}$ & $SU(4)_{(y_1,y_2,y_3)}$ & $U(1)_R$\\
\hline\hline
$A_{11}=P_{12}$ & $p_1 p_5$ & $\left(+1,~0,~0\right)$ & $\left(+1,~0,~0\right)$ & $r_1+r_5$\\
$A_{12}=P_{23}$ & $p_1 p_6$ & $\left(+1,~0,~0\right)$ & $\left(-1,+1,0\right)$ & $r_1+r_6$\\
$A_{13}=P_{34}$ & $p_1 p_7$ & $\left(+1,~0,~0\right)$ & $\left(0,-1,+1\right)$ & $r_1+r_7$\\
$A_{14}=P_{41}$ & $p_1 p_8$ & $\left(+1,~0,~0\right)$ & $\left(0,~~0,-1\right)$ & $r_1+r_8$\\
\hline
$A_{21}=X_{12}$ & $p_2  p_5$ & $\left(-1,+1,0\right)$ & $\left(+1,~0,~0\right)$ & $r_2+r_5$\\
$A_{22}=X_{23}$ & $p_2  p_6$ & $\left(-1,+1,0\right)$ & $\left(-1,+1,0\right)$  & $r_2+r_6$\\
$A_{23}=X_{34}$ & $p_2  p_7$ & $\left(-1,+1,0\right)$ & $\left(0,-1,+1\right)$  & $r_2+r_7$\\
$A_{24}=X_{41}$ & $p_2  p_8$ & $\left(-1,+1,0\right)$ & $\left(0,~~0,-1\right)$ & $r_2+r_8$\\
\hline
$A_{31}=Y_{12}$ & $p_3  p_5$ & $\left(0,-1,+1\right)$ & $\left(+1,~0,~0\right)$ & $r_3+r_5$\\
$A_{32}=Y_{23}$ & $p_3  p_6$ & $\left(0,-1,+1\right)$ & $\left(-1,+1,0\right)$  & $r_3+r_6$\\
$A_{33}=Y_{34}$ & $p_3  p_7$ & $\left(0,-1,+1\right)$ & $\left(0,-1,+1\right)$  & $r_3+r_7$\\
$A_{34}=Y_{41}$ & $p_3  p_8$ & $\left(0,-1,+1\right)$ & $\left(0,~~0,-1\right)$ & $r_3+r_8$\\
\hline
$A_{41}=Z_{12}$ & $p_4  p_5$ & $\left(0,~~0,-1\right)$ & $\left(+1,~0,~0\right)$ & $r_4+r_5$\\
$A_{42}=Z_{23}$ & $p_4  p_6$ & $\left(0,~~0,-1\right)$ & $\left(-1,+1,0\right)$  & $r_4+r_6$\\
$A_{43}=Z_{34}$ & $p_4  p_7$ & $\left(0,~~0,-1\right)$ & $\left(0,-1,+1\right)$  & $r_4+r_7$\\
$A_{44}=Z_{41}$ & $p_4  p_8$ & $\left(0,~~0,-1\right)$ & $\left(0,~~0,-1\right)$ & $r_4+r_8$\\
\hline
\end{tabular}}
\caption{
Generators of the master space ${}^{\text{Irr}}\mathcal{F}^\flat$ for the $\mathbb{C}^4/\mathbb{Z}_4 ~(1,1,1,1)$ with the global symmetry charges. 
\label{tex04}
}
\end{table}
%--------------------------------

%======================================================================
\subsection{The Master Space for $Q^{1,1,1}$ \label{sec:043}}
%======================================================================

%--------------------------------
\begin{figure}[H]
\begin{center}
\resizebox{0.3\hsize}{!}{
\includegraphics[height=6cm]{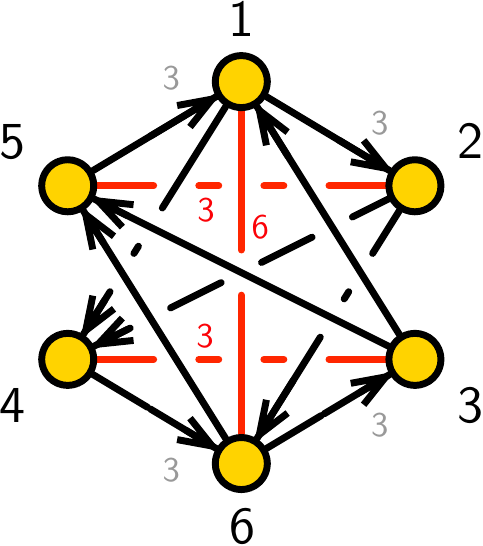} 
}
\caption{
The quiver diagram for the $Q^{1,1,1}$ model. 
\label{f12a01}}
\end{center}
\end{figure}
%--------------------------------

The $J$- and $E$-terms of the $Q^{1,1,1}$ model \cite{DAuria:1983sda, 1986PhR...130....1D,Nilsson:1984bj,Franco:2015tna,Franco:2015tya} are as follows,
\beal{es12a01}
\resizebox{0.95\textwidth}{!}{$
\begin{array}{rrclrcl}
 &  & J &  & & E &  \\
\Lambda_{21}^{1} : & X_{12}\cdot Y_{24}\cdot Y_{41}\cdot Y_{12} &-& Y_{12}\cdot Y_{23}\cdot Y_{31}\cdot X_{12} 
& X_{23}\cdot X_{31} &-& X_{24}\cdot X_{41} \\
\Lambda_{21}^{2}: & Y_{12}\cdot X_{24}\cdot Y_{41}\cdot X_{12} &-& X_{12}\cdot Y_{23}\cdot X_{31}\cdot Y_{12} 
&X_{23}\cdot Y_{31} &-& Y_{24}\cdot X_{41} \\
\Lambda_{21}^{3} : & X_{12}\cdot Y_{24}\cdot X_{41}\cdot Y_{12} &-& Y_{12}\cdot X_{23}\cdot Y_{31}\cdot X_{12} 
& X_{24}\cdot Y_{41} &-& Y_{23}\cdot X_{31} \\
\Lambda_{21}^{4} : & X_{12}\cdot X_{23}\cdot X_{31}\cdot Y_{12} &- &Y_{12}\cdot X_{24}\cdot X_{41}\cdot X_{12} 
&Y_{23}\cdot Y_{31} &-& Y_{24}\cdot Y_{41} \\
\Lambda_{43} : & Y_{31}\cdot Y_{12} \cdot X_{24} &-& X_{31}\cdot Y_{12}\cdot Y_{24} 
& X_{41}\cdot X_{12}\cdot Y_{23} &-& Y_{41}\cdot X_{12}\cdot X_{23} \\
\Lambda_{34} : & Y_{41}\cdot Y_{12}\cdot X_{23} &-& X_{41}\cdot Y_{12}\cdot Y_{23} 
& X_{31}\cdot X_{12}\cdot Y_{24} &-& Y_{31}\cdot X_{12}\cdot X_{24} 
\end{array} 
~,~
$}
\nn\\
\eea
where \fref{f12a01} shows the corresponding quiver diagram.
The $J$- and $E$-terms can be rewritten in terms of $G+3=7$ independent new variables, which are
\beal{es12a01a} 
v_1 = X_{12} ~,~
v_2 = Y_{12} ~,~
v_3 = X_{23} ~,~
v_4 = X_{31} ~,~
v_5 = Y_{31} ~,~
v_6 = X_{41} ~,~
v_7 = Y_{41} ~.~
\nn\\
\eea
These independent fields under the $J$- and $E$-terms are encoded in the $K$-matrix, which takes the form
\beal{es12a01b}
K= 
\resizebox{0.285\textwidth}{!}{$
\left(
\begin{array}{c|ccccccc}
\; & v_1 & v_2 & v_3 & v_4 & v_5 & v_6& v_7
\\
\hline
X_{12}& 1 & 0 & 0 & 0 & 0 & 0 & 0 \\
Y_{12}& 0 & 1 & 0 & 0 & 0 & 0 & 0 \\
X_{31}& 0 & 0 & 0 & 1 & 0 & 0 & 0 \\
Y_{31}& 0 & 0 & 0 & 0 & 1 & 0 & 0 \\
X_{41}& 0 & 0 & 0 & 0 & 0 & 1 & 0 \\
Y_{41}& 0 & 0 & 0 & 0 & 0 & 0 & 1 \\
X_{24}& 0 & 0 & 1 & 1 & 0 & -1 & 0 \\
Y_{24}& 0 & 0 & 1 & 0 & 1 & -1 & 0 \\
X_{23}& 0 & 0 & 1 & 0 & 0 & 0 & 0 \\
Y_{23}& 0 & 0 & 1 & 0 & 0 & -1 & 1 \\
\end{array}
\right)
$}
~.~
\eea
Following the forward algorithm, we obtain the $P$-matrix, which takes the following form
\beal{es12a02}
P=
\resizebox{0.3\textwidth}{!}{$
\left(
\begin{array}{c|cc|cccc|c|c}
\; & p_1 & p_2 & p_3 & p_4 & p_5 & p_6 & p_7 & o  \\
\hline
X_{12}& 1 & 0 & 0 & 0 & 0 & 0 &0  &0\\
Y_{12}& 0 & 1 & 0 & 0 & 0 & 0 &0  &0\\
X_{31}& 0 & 0 & 1 & 0 & 0 & 0 &1  &0\\
Y_{31}& 0 & 0 & 0 & 1 & 0 & 0 &1  &0\\
X_{41}& 0 & 0 & 0 & 0 & 1 & 0 &1  &0\\
Y_{41}& 0 & 0 & 0 & 0 & 0 & 1 &1  &0\\
X_{24}& 0 & 0 & 1 & 0 & 0 & 0 &0  &1\\
Y_{24}& 0 & 0 & 0 & 1 & 0 & 0 &0  &1\\
X_{23}& 0 & 0 & 0 & 0 & 1 & 0 &0  &1\\
Y_{23}& 0 & 0 & 0 & 0 & 0 & 1 &0  &1\\
\end{array}
\right) 
$}
~.~
\eea
The $P$-matrix defines the GLSM fields $p_1,\dots,p_7$ in terms of the chiral fields
of the $Q^{1,1,1}$ model. 
We also have an extra GLSM field, which we call $o$.
Note that, in comparison to the mesonic moduli space of the $Q^{1,1,1}$ brane brick model \cite{Franco:2015tna, Franco:2015tya}, which has two extra GLSM fields,
we identify for the master space ${}^{\text{Irr}}\mathcal{F}^\flat$ only a single extra GLSM field. 
We will see later why we only have a single extra GLSM field for the master space ${}^{\text{Irr}}\mathcal{F}^\flat$ when we look at its toric diagram.

In order to obtain the toric diagram of the master space ${}^{\text{Irr}}\mathcal{F}^\flat$, 
we identify the $U(1)$ charges due to the $J$- and $E$-terms on the GLSM fields of the brane brick model.
These charges are summarized in the following $Q_{JE}$-charge matrix,
\beal{es12a03}
Q_{JE}=
\resizebox{0.285\textwidth}{!}{$
\left(
\begin{array}{cc|cccc|c|c}
p_1 & p_2 & p_3 & p_4 & p_5 & p_6 & p_7 & o  \\
\hline
0 & 0 & -1 & -1 & -1 & -1 & 1 & 1 
\end{array}
\right)
$}
~.~
\eea
Additionally, the GLSM fields carry $U(1)$ charges due to the $D$-terms of the brane brick model.
These charges are summarized in the following $Q_D$-matrix,
\beal{es12a03b}
Q_D=
\resizebox{0.255\textwidth}{!}{$
\left(
\begin{array}{cc|cccc|c|c}
p_1 & p_2 & p_3 & p_4 & p_5 & p_6 & p_7 & o  \\
\hline
 1 & 1 & 0 & 0 & 0 & 0 & -1 & 0 \\
 0 & 0 & 1 & 1 & 0 & 0 & -1 & 0 \\
 0 & 0 & 0 & 0 & 1 & 1 & -1 & 0 \\
\end{array}
\right)
$}
~.~
\eea
Combined, the $Q_{JE}$- and $Q_{D}$-matrices indicate that the pairs of GLSM fields $(p_1,p_2)$, $(p_3,p_4)$ and $(p_5,p_6)$ carry the same $U(1)$ charges under the $J$-, $E$- and $D$-terms of the brane brick model.
We note from this that the global symmetry of the mesonic moduli space $\mathcal{M}^{mes}$ of the brane brick model is enhanced to 
\beal{es12a04a}
SU(2)_{x} \times SU(2)_{y} \times SU(2)_{z} \times U(1)_R
~,~
\eea
with 3 $SU(2)$ factors each corresponding to a pair of GLSM fields carrying the same $Q_{JE}$- and $Q_{D}$-charges.
When we focus only on the charges given by the $Q_{JE}$-matrix, we note that the GLSM fields $(p_1,p_2)$ and $(p_3,p_4,p_5,p_6)$ carry the same charges. 
This indicates that the global symmetry for the master space ${}^{\text{Irr}}\mathcal{F}^\flat$ is enhanced to 
\beal{es12a04}
SU(4)_{(x_1,x_2,x_3)} \times SU(2)_y \times U(1)_{b_1} \times U(1)_{b_2} \times U(1)_R ~,~
\eea
where the total rank of the global symmetry of the master space ${}^{\text{Irr}}\mathcal{F}^\flat$ is as expected $G+3=7$. 
\tref{tex05} summarizes how the GLSM fields of the $Q^{1,1,1}$ model are charged under the global symmetry of the master space ${}^{\text{Irr}}\mathcal{F}^\flat$.

%--------------------------------
\begin{table}[ht!]
\centering
\begin{tabular}{|c|c|c|c|c|c|c|}
\hline
\; & $SU(4)_{(x_1,x_2,x_3)}$ & $SU(2)_{y}$ &$U(1)_{b_1}$& $U(1)_{b_2}$ & $U(1)_{R}$ & fugacity\\
\hline\hline
$p_1$ & $0$ & $+1$ & $-2$ &$+1$&$r_1$ &$t_1 = y b_1^{-2} b_2 t$  \\
$p_2$ & $0$ & $-1$ & $-2$  &$+1$&$r_2$ &$t_2 = y^{-1} b_1^{-2} b_2 t$  \\
\hline
$p_3$ & $(+1,~0,~0)$ & $0$    & $+1$ & $0$  & $r_3$ &    $t_3 = x_1 b_1 t $  \\
$p_4$ & $(-1,+1,0)$    & $0$    & $+1$ & $0$  & $r_4$ &    $t_4 = x_1^{-1} x_2 b_1 t$  \\
$p_5$ & $(0,-1,+1)$    & $0$    & $+1$ & $0$  & $r_5$ &    $t_5=  x_2^{-1} x_3 b_1 t$  \\
$p_6$ & $(0,~~0,-1)$  & $0$    & $+1$ & $0$  & $r_6$ &    $t_6 = x_3^{-1} b_1 t$  \\
\hline
$p_7$ & $0$ & $0$  & $0$ &$-2$ & $r_7$ &    $t_7= b_2^{-2} t$  \\
\hline
$o$ & $0$ & $0$  & $0$ &$0$ & $0$ &    $ u = 1$  \\
\hline
\end{tabular}
\caption{The global symmetry charges of the master space ${}^{\text{Irr}}\mathcal{F}^\flat$ on the GLSM fields $p_\alpha$ for the $Q^{1,1,1}$ model.
\label{tex05}}
\end{table}
%--------------------------------

Given that the $J$- and $E$-terms in \eref{es12a01} are all binomial, the master space ${}^{\text{Irr}}\mathcal{F}^\flat$ is toric and has a $6$-dimensional toric diagram which is given by
\beal{es12a05}
G_t^{{}^{\text{Irr}}\mathcal{F}^\flat}=
\resizebox{0.255\textwidth}{!}{$
\left(
\begin{array}{cc|cccc|c|c}
p_1 & p_2 & p_3 & p_4 & p_5 & p_6 & p_7 & o  \\
\hline
 1 & 0 & 0 & 0 & 0 & 0 & 0 & 1 \\
 0 & 1 & 0 & 0 & 0 & 0 & 0 & 1 \\
 0 & 0 & 1 & 0 & 0 & 0 & 0 & 1 \\
 0 & 0 & 0 & 1 & 0 & 0 & 0 & 1 \\
 0 & 0 & 0 & 0 & 1 & 0 & 0 & 0 \\
 0 & 0 & 0 & 0 & 0 & 1 & 0 & 0 \\
 \hline
 1 & 1 & 1 & 1 & 1 & 1 & 1 & 3 \\
\end{array}
\right)
$}
~.~
\eea
We note here that the toric diagram, a convex polytope on a 6-dimensional hyperplane is made of 7 extremal vertices corresponding each to one of the GLSM fields $p_1,\dots,p_7$.
The extra GLSM field $o$ corresponds to a vertex which lies outside the 6-dimensional hyperplane in $\mathbb{Z}^7$. 
It is not part of the toric diagram of the master space ${}^{\text{Irr}}\mathcal{F}^\flat$ of the $Q^{1,1,1}$ brane brick model and we can identify the extra GLSM field $o$ as an over-parameterization of the master space ${}^{\text{Irr}}\mathcal{F}^\flat$.
In other words, when we identify the generators and defining relations of the master space ${}^{\text{Irr}}\mathcal{F}^\flat$ in terms of GLSM fields, the presence or absence of the extra GLSM $o$ does not affect the shape and number of generators and defining relations of the master space ${}^{\text{Irr}}\mathcal{F}^\flat$. 

This over-parameterization of the master space ${}^{\text{Irr}}\mathcal{F}^\flat$ by the extra GLSM field $o$ is best observed when we calculate the Hilbert series of ${}^{\text{Irr}}\mathcal{F}^\flat$ in terms of fugacities that count degrees in GLSM fields. 
We can calculate the Hilbert series of the master space ${}^{\text{Irr}}\mathcal{F}^\flat$  in terms of fugacities corresponding to GLSM fields by using the symplectic quotient description of ${}^{\text{Irr}}\mathcal{F}^\flat$ and the corresponding Molien integral formula for the Hilbert series in \eref{es06a20}. 
Accordingly, the Hilbert series for ${}^{\text{Irr}}\mathcal{F}^\flat$ takes the form
\beal{es12a20}
&&
g(
t_\alpha,u;
{}^{\text{Irr}}\mathcal{F}^\flat
)
=
(1-u t_3 t_4 t_7-u t_3 t_5 t_7-u t_3 t_6 t_7-u t_4 t_5 t_7-u t_4 t_6 t_7-u t_5 t_6 t_7
\nn\\
&&
\hspace{1cm}
+u^2 t_3 t_4 t_5 t_7+u^2 t_3 t_4 t_6 t_7+u^2 t_3 t_5 t_6 t_7+u^2 t_4 t_5 t_6 t_7+u t_3 t_4 t_5 t_7^2+u t_3 t_4 t_6 t_7^2
\nn\\
&&
\hspace{1cm}
+u t_3 t_5 t_6 t_7^2+u t_4 t_5 t_6 t_7^2-u t_3  t_4 t_5 t_6 t_7^3-u^2 t_3 t_4 t_5 t_6 t_7^2 -u^3 t_3 t_4 t_5 t_6 t_7
)
\nn\\
&&
\hspace{1cm}
\times\frac{1}{(1-t_1) (1-t_2) (1-u t_3) (1-u t_4) (1-u t_5) (1-ut_6)(1-t_3 t_7) (1-t_4 t_7) }
\nn\\
&&
\hspace{1cm}
\times \frac{1}{(1-t_5 t_7) (1-t_6 t_7)}
~,~
\eea
where the fugacities $t_\alpha$ correspond to the GLSM fields $p_\alpha$ and the fugacity $u$ corresponds to the extra GLSM field $o$.
Even if we set the fugacity $u=1$, the Hilbert series in \eref{es12a20} describes the same master space ${}^{\text{Irr}}\mathcal{F}^\flat$ for the $Q^{1,1,1}$ brane brick model.
This is because, when we calculate the corresponding plethystic logarithm of the Hilbert series,
\beal{es12a22}
&&
\text{PL}\left[
g(
t_\alpha,u;
{}^{\text{Irr}}\mathcal{F}^\flat
)
\right]
= t_1 + t_2 +u t_3+u t_4+u t_5+u t_6+t_3 t_7 +t_4 t_7+t_5 t_7+t_6 t_7
\nn\\
&&
\hspace{1cm}
-(u t_3 t_4 t_7+u t_3 t_5 t_7+u t_3 t_6 t_7+u t_4 t_5 t_7+u t_4 t_6 t_7+u t_5
   t_6 t_7)
   +\dots
 ~,~
\eea
the negative terms corresponding to first order generator relations describe the same first order relations between the generators with or without the fugacity $u$ counting the degree in the extra GLSM field $o$.
The plethystic logarithm also indicates that the master space ${}^{\text{Irr}}\mathcal{F}^\flat$
here is a non-complete intersection.

Using the $P$-matrix, we can express the chiral fields of the brane brick model as products of GLSM fields as follows, 
\beal{es12a30}
&
X_{12} = p_1 ~,~
Y_{12} = p_2 ~,~
X_{31} = p_3 p_7 ~,~
Y_{31} = p_4 p_7 ~,~
X_{41} = p_5 p_7 ~,~
Y_{41} = p_6 p_7 ~,~
&
\nn\\
&
X_{24} = p_3 o ~,~
Y_{24} = p_4 o ~,~
X_{23} = p_5 o ~,~
Y_{23} = p_6 o ~.~
&
~.~
\eea
Using primary decomposition of the $J$- and $E$-terms in \eref{es12a01}, we obtain the coherent component as follows,
\beal{es12a35}
&&
\mathcal{I}_{JE}^{\text{Irr}} =
\big\langle
~
X_{24} Y_{41}-Y_{23} X_{31}~,~
Y_{24} X_{41}- X_{23} Y_{31} ~,~
\nn\\
&&
\hspace{1.5cm}
X_{24} X_{41}-X_{23} X_{31}~,~
Y_{24} Y_{41}-Y_{23} Y_{31} ~,~
\nn\\
&&
\hspace{1.5cm}
X_{23} Y_{41}- Y_{23} X_{41}~,~
X_{24} Y_{31}- Y_{24} X_{31}
~
\big\rangle
~.~
\eea
The master space ${}^{\text{Irr}}\mathcal{F}^\flat$ is then given by the following quotient,
\beal{es12a36}
{}^{\text{Irr}}\mathcal{F}^\flat
= 
\text{Spec}~
\mathbb{C}^{10}[
X_{ij},
Y_{ij}
]
/ \mathcal{I}_{JE}^{\text{Irr}}
~.~
\eea
Under the following fugacity map,
\beal{es12a40}
&
t = t_1^{1/7} t_2^{1/7} t_3^{1/7} t_4^{1/7} t_5^{1/7} t_6^{1/7} t_7^{1/7} u^{3/7}~,~
x_1 = \frac{t_3^{3/4}}{t_4^{1/4} t_5^{1/4} t_6^{1/4}}~,~ 
x_2 = \frac{t_3^{1/2} t_4^{1/2}}{t_5^{1/2} t_6^{1/2}}~,~ 
x_3 = \frac{t_3^{1/4} t_4^{1/4} t_5^{1/4}}{t_6^{3/4}}~,~ 
&
\nn\\
&
y = \frac{t_1^{1/2}}{t_2^{1/2}}~,~
b_1 = \frac{t_3^{3/28} t_4^{3/28} t_5^{3/28} t_6^{3/28} u^{4/7}} {t_1^{1/7} t_2^{1/7} t_3^{1/7}}~,~
b_2 = \frac{t_1^{1/14} t_2^{1/14} t_3^{1/14} t_4^{1/14} t_5^{1/14} t_6^{1/14} u^{5/7}}{t_7^{3/7}}~,~ 
&
\eea
we can rewrite the refined Hilbert series in \eref{es12a20} in terms of characters of irreducible representations of the global symmetry of the master space ${}^{\text{Irr}}\mathcal{F}^\flat$.
Note that here again, we can set the fugacity for the extra GLSM field to $u=1$ without loss of generality. 

The refined Hilbert series in terms of characters of irreducible representations of the global symmetry of the master space  ${}^{\text{Irr}}\mathcal{F}^\flat$ takes the following form,
\beal{es12a41}
g (t,x_i,y,b_j;{}^{\text{Irr}}\mathcal{F}^\flat )
=\sum_{n_1,n_2,n_3=0}^\infty [n_1+n_2,0,0;n_3] b_1^{n_1+n_2-2 n_3} b_2^{n_3-2 n_1} t^{n_1+2n_2+n_3}
~,~
\eea
where $[m_1,m_2,m_3;n]=[m_1,m_2,m_3]_{SU(4)_{(x_1,x_2,x_3)}} [n]_{SU(2)_{y}}$.
We can write the character expansion in \eref{es12a41} as a highest weight generating function \cite{Hanany:2014dia}.
The corresponding highest weight generating function takes the form,
\beal{es12a42}
h\left(t,\mu_i,\nu,b_j;{}^{\text{Irr}}\mathcal{F}^\flat\right)&=&\frac{1}{\left(1-\nu b_1^{-2} b_2 t \right) \left(1- \mu_1 b_1 t \right)  \left(1- \mu_1 b_1 b_2^{-2} t^2 \right) }
\eea
where $ \mu_1^{m_1} \mu_2^{m_2} \mu_3^{m_3} \nu^{n}  \sim [m_1,m_2,m_3]_{SU(4)_{(x_1,x_2,x_3)}} [n]_{SU(2)_{y}}$.
The plethystic logarithm of the refined Hilbert series in terms of characters of irreducible representations of the master space global symmetry takes the following form,
\beal{es12a45}
&&
PL[g (t,x_i,y,b_j;{}^{\text{Irr}}\mathcal{F}^\flat )]
= 
[0,0,0;1] b_1^{-2} b_2 t+[1,0,0;0] b_1t+[1,0,0;0] b_1 b_2^{-2} t^2
\nn\\
&&
\hspace{1cm}
-[0,1,0;0] b_1^2 b_2^{-2} t^3 +\dots ~.~ 
\eea
From the plethystic logarithm, we identify the generators of the master space ${}^{\text{Irr}}\mathcal{F}^\flat$ as,
\beal{es12a46}
A_{i} =(X_{12},Y_{12})_i
&~~\leftrightarrow~~&
+[0,0,0;1] b_1^{-2} b_2 t
~.~
\\
B_{j} =(X_{31},Y_{31},X_{41},Y_{41})_j
&~~\leftrightarrow~~&
+[1,0,0;0] b_1t
~.~
\\
C_{k} =(X_{24},Y_{24},X_{23},Y_{23})_k
&~~\leftrightarrow~~&
+[1,0,0;0] b_1 b_2^{-2} t^2
~.~
\eea
The relation at order $-[0,1,0;0] b_1^2 b_2^{-2} t^3$ is given by,
\beal{es12a47}
N^{lm}=\epsilon^{lmjk} B_{j} C_{k}=0
 &~~\leftrightarrow~~& -[0,1,0;0] b_1^2 b_2^{-2} t^3
~,~
\eea
where $N^{lm}=-N^{ml}$.
As noted above, the presence or absence of the extra GLSM field $o$ does not affect the algebraic description of the generators and first order relations of the master space ${}^{\text{Irr}}\mathcal{F}^\flat$ of the $Q^{1,1,1}$ brane brick model.
\tref{tex06} summarizes the generators of the master space ${}^{\text{Irr}}\mathcal{F}^\flat$ with their global symmetry charges.

%--------------------------------
\begin{table}[ht!]
\centering
\resizebox{0.9\textwidth}{!}{
\begin{tabular}{|c|c|c|c|c|c|c|c|}
\hline
generators  & GLSM fields & $SU(4)_{(x_1,x_2,x_3)}$ & $SU(2)_{y}$ & $U(1)_{b_1}$ & $U(1)_{b_2}$ & $U(1)_R$\\
\hline\hline
$A_1=X_{12}$ & $p_1 $ & $0$ &$+1$&$-2$& $+1$ & $r_1$\\
$A_2=Y_{12}$ & $p_2 $ & $0$ &$-1$&$-2$&$+1$ & $r_2$\\
\hline
$B_{11}=X_{31}$ & $p_3 p_7 $ & $\left(+1,~0,~0\right)$ &$0$&$+1$&$0$ & $r_3+r_7$\\
$B_{22}=Y_{31}$ & $p_4 p_7 $ & $\left(-1,+1,0\right)$ &$0$&$+1$&$0$ & $r_4+r_7$\\
$B_{12}=X_{41}$ & $p_5 p_7 $ & $\left(0,-1,+1\right)$ &$0$&$+1$&$0$  & $r_5+r_7$\\
$B_{21}=Y_{41}$ & $p_6 p_7 $ & $\left(0,~~0,-1\right)$ &$0$&$+1$&$0$  & $r_6+r_7$\\
\hline
$C_{11}=X_{24}$ & $p_3 o $ & $\left(+1,~0,~0\right)$ &$0$&$+1$&$-2$ & $r_3$\\
$C_{22}=Y_{24}$ & $p_4 o $ & $\left(-1,+1,0\right)$ &$0$&$+1$&$-2$ & $r_4$\\
$C_{12}=X_{23}$ & $p_5 o $ & $\left(0,-1,+1\right)$ &$0$&$+1$&$-2$  & $r_5$\\
$C_{21}=Y_{23}$ & $p_6 o $ & $\left(0,~~0,-1\right)$ &$0$&$+1$&$-2$  & $r_6$\\
\hline
\end{tabular}}
\caption{
Generators of the master space ${}^{\text{Irr}}\mathcal{F}^\flat$ for the $Q^{1,1,1}$ model with the global symmetry charges. 
\label{tex06}
}
\end{table}
%--------------------------------

We have in this section identified the master space ${}^{\text{Irr}}\mathcal{F}^\flat$ of the $Q^{1,1,1}$ brane brick model to be a 7-dimensional affine toric variety. 
However, although so far we have encountered master spaces ${}^{\text{Irr}}\mathcal{F}^\flat$ for brane brick models which were toric and Calabi-Yau, the master space ${}^{\text{Irr}}\mathcal{F}^\flat$ of the $Q^{1,1,1}$ brane brick model appears to be toric but \textit{not} Calabi-Yau.
This phenomenon can be seen when we unrefine the Hilbert series of the master space ${}^{\text{Irr}}\mathcal{F}^\flat$ of the $Q^{1,1,1}$ brane brick model in \eref{es12a20} by setting the fugacities $t_\alpha = t$ and $u=1$.
This results in the unrefined Hilbert series of the form
\beal{es12a21}
g(
t;
{}^{\text{Irr}}\mathcal{F}^\flat
)
=
\frac{1 - 6 t^3 + 4t^4+3t^5-t^6-t^7}{
(1 - t)^6 (1 - t^2)^4
} ~,~
\eea
where we discover that the numerator of the unrefined Hilbert series is \textit{not} palindromic. 
By Stanley's theorem \cite{stanley1978hilbert},
the numerator of the Hilbert series in rational form is palindromic if the corresponding coordinate ring is Gorenstein and the variety is Calabi-Yau. 
While the coherent component of the $J$- and $E$-terms of the $Q^{1,1,1}$ brane brick model is binomial, implying that the master space ${}^{\text{Irr}}\mathcal{F}^\flat$ is toric \cite{fulton,cox1995homogeneous}, the non-palindromic numerator of the unrefined Hilbert series in \eref{es12a21} indicates that the master space ${}^{\text{Irr}}\mathcal{F}^\flat$ is indeed not Calabi-Yau.
\\

%======================================================================
\subsection{The Master Space for $Y^{2,4}(\mathbb{CP}^2)$ \label{sec:044}}
%======================================================================

\begin{figure}[H]
\begin{center}
\resizebox{0.3\hsize}{!}{
\includegraphics[height=6cm]{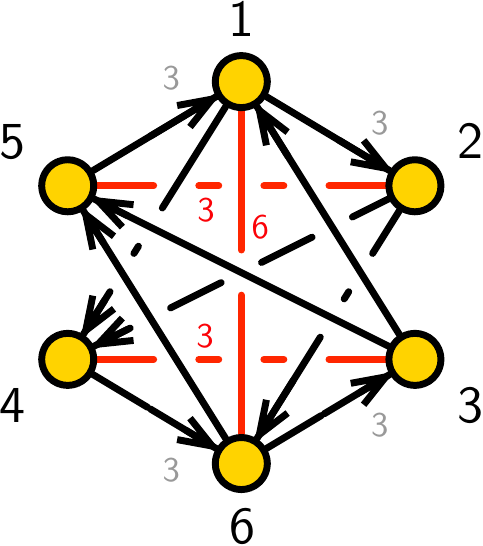} 
}
\caption{
The quiver diagram for the $Y^{2,4}(\mathbb{CP}^2)$ model. 
\label{f13a01}}
\end{center}
\end{figure}

The quiver diagram for the $Y^{2,4}(\mathbb{CP}^2)$ model \cite{Martelli:2008rt, Gauntlett:2004hh,Franco:2022isw} is shown in \fref{f13a01}.
The corresponding $J$- and $E$-terms take the following form,
\beal{es13a01}
\begin{array}{rrclcrcl}
 &  & J &  & & & E &  \\
\Lambda_{16}^{1}  : & X_{63} \cdot P_{35} \cdot  Y_{51}  &-&   Y_{63} \cdot P_{35} \cdot X_{51}   
&~~&    Z_{12} \cdot Q_{26}  &-&  Q_{14} \cdot Z_{46} \\ 
\Lambda_{16}^{2}  : & Y_{63} \cdot P_{35} \cdot Z_{51}  &-&  Z_{63} \cdot P_{35} \cdot  Y_{51}
&~~ &     X_{12} \cdot Q_{26} &-&  Q_{14} \cdot X_{46}  \\ 
\Lambda_{16}^{3}  : & Z_{63} \cdot P_{35} \cdot  X_{51}  &-&   X_{63} \cdot P_{35} \cdot Z_{51}   
&~~ &    Y_{12} \cdot Q_{26}  &-& Q_{14} \cdot Y_{46} \\ 
\Lambda_{61}^{1}  : & X_{12} \cdot P_{24} \cdot  Y_{46}  &-&  Y_{12} \cdot  P_{24} \cdot X_{46}   
&~~ &    Z_{63} \cdot Q_{31}  &-&  Q_{65} \cdot Z_{51} \\ 
\Lambda_{61}^{2}  : & Y_{12} \cdot P_{24} \cdot  Z_{46}  &-&  Z_{12} \cdot  P_{24} \cdot Y_{46}   
&~~ &    X_{63} \cdot Q_{31}  &-& Q_{65} \cdot X _{51}  \\ 
\Lambda_{61}^{3}  : & Z_{12} \cdot P_{24} \cdot  X_{46}  &-&  X_{12} \cdot  P_{24} \cdot Z_{46}   
&~~ &    Y_{63} \cdot Q_{31}  &-& Q_{65} \cdot Y_{51} \\
\Lambda_{43}^{1}  : & P_{35} \cdot X_{51} \cdot  Q_{14}  &-&  Q_{31} \cdot  X_{12} \cdot  P_{24}   
&~~ &    Y_{46} \cdot  Z_{63}  &-& Z_{46} \cdot Y_{63} \\ 
\Lambda_{43}^{2}  : & P_{35} \cdot Y_{51} \cdot  Q_{14}  &-&  Q_{31} \cdot  Y_{12} \cdot  P_{24}   
&~~ &    Z_{46} \cdot  X_{63}  &-&  X_{46} \cdot Z_{63} \\ 
\Lambda_{43}^{3}  : & P_{35} \cdot Z_{51} \cdot  Q_{14}  &-&  Q_{31} \cdot  Z_{12} \cdot  P_{24}   
&~~ &    X_{46} \cdot  Y_{63}  &-& Y_{46} \cdot X_{63} \\
\Lambda_{52}^{1}  : & P_{24} \cdot X_{46} \cdot  Q_{65}  &-&  Q_{26} \cdot  X_{63} \cdot  P_{35}   
&~~ &    Y_{51} \cdot Z_{12}  &-& Z_{51} \cdot Y_{12} \\  
\Lambda_{52}^{2}  : & P_{24} \cdot Y_{46} \cdot  Q_{65}  &-&  Q_{26} \cdot  Y_{63} \cdot  P_{35}   
&~~ &    Z_{51} \cdot X_{12}  &-& X_{51} \cdot Z_{12} \\ 
\Lambda_{52}^{3}  : & P_{24} \cdot Z_{46} \cdot  Q_{65}  &-&  Q_{26} \cdot  Z_{63} \cdot  P_{35}   
&~~ &    X_{51} \cdot  Y_{12}  &-&  Y_{51} \cdot X_{12}  
 \end{array} 
~.~
\eea
We can rewrite the $J$- and $E$-terms in terms of $G+3=9$ independent new variables, which are
\beal{es13a01a} 
&
v_1 = P_{24} ~,~
v_2 = P_{35} ~,~
v_3 = Q_{14} ~,~
v_4 = Q_{31} ~,~
v_5 = Q_{65} ~,~
v_6 = Q_{26} ~,~
&
\nn\\
&
v_7 = Z_{46} ~,~
v_8 = X_{51} ~,~
v_9 = Y_{46} ~.~
\eea
These independent fields are related to the rest of the chiral fields in the $Y^{2,4}(\mathbb{CP}^2)$ model. This relationship is encoded in the following $K$-matrix,
\beal{es13a01b}
K= 
\resizebox{0.37\textwidth}{!}{$
\left(
\begin{array}{c|ccccccccc}
\; & v_1 & v_2 & v_3 & v_4 & v_5 & v_6 & v_7 & v_8 & v_9 
\\
\hline
P_ {35}& 0 & 1 & 0 & 0 & 0 & 0 & 0 & 0 & 0 \\
P_ {24}& 1 & 0 & 0 & 0 & 0 & 0 & 0 & 0 & 0 \\
X_ {46}& -1 & 1 & 0 & -1 & 0 & 1 & 0 & 1 & 0 \\
Y_ {46}& 0 & 0 & 0 & 0 & 0 & 0 & 0 & 0 & 1 \\
Z_ {46}& 0 & 0 & 0 & 0 & 0 & 0 & 1 & 0 & 0 \\
X_ {12}& -1 & 1 & 1 & -1 & 0 & 0 & 0 & 1 & 0 \\
Y_ {12}& 0 & 0 & 1 & 0 & 0 & -1 & 0 & 0 & 1 \\
Z_ {12}& 0 & 0 & 1 & 0 & 0 & -1 & 1 & 0 & 0 \\
X_ {51}& 0 & 0 & 0 & 0 & 0 & 0 & 0 & 1 & 0 \\
Y_ {51}& 1 & -1 & 0 & 1 & 0 & -1 & 0 & 0 & 1 \\
Z_ {51}& 1 & -1 & 0 & 1 & 0 & -1 & 1 & 0 & 0 \\
X_ {63}& 0 & 0 & 0 & -1 & 1 & 0 & 0 & 1 & 0 \\
Y_ {63}& 1 & -1 & 0 & 0 & 1 & -1 & 0 & 0 & 1 \\
Z_ {63}& 1 & -1 & 0 & 0 & 1 & -1 & 1 & 0 & 0 \\
Q_ {26}& 0 & 0 & 0 & 0 & 0 & 1 & 0 & 0 & 0 \\
Q_ {14}& 0 & 0 & 1 & 0 & 0 & 0 & 0 & 0 & 0 \\
Q_ {31}& 0 & 0 & 0 & 1 & 0 & 0 & 0 & 0 & 0 \\
Q_ {65}& 0 & 0 & 0 & 0 & 1 & 0 & 0 & 0 & 0 \\
\end{array}
\right)
$}
~.~
\eea

%--------------------------------
%--------------------------------

\noindent
Using the forward algorithm, we obtain the $P$-matrix as follows,
\beal{es13a02}
P=
\resizebox{0.5\textwidth}{!}{$
\left(
\begin{array}{c|ccc|cc|cc|cc|cc|cc}
\; & p_1 & p_2 & p_3 & p_4 & p_5 & p_6 & p_7 & p_8 & p_9 & p_{10} & p_{11}  & o_1 & o_2\\
\hline
P_ {35} & 0 & 0 & 0 & 1 & 0 & 0 & 0 & 0 & 0 & 0 & 1 & 1 & 0 \\
P_ {24} &0 & 0 & 0 & 1 & 0 & 0 & 0 & 0 & 0 & 1 & 0 & 0 & 1 \\
X_ {46} & 1 & 0 & 0 & 0 & 0 & 1 & 0 & 0 & 0 & 0 & 0 & 1 & 0 \\
Y_ {46} & 0 & 1 & 0 & 0 & 0 & 1 & 0 & 0 & 0 & 0 & 0 & 1 & 0 \\
Z_ {46} & 0 & 0 & 1 & 0 & 0 & 1 & 0 & 0 & 0 & 0 & 0 & 1 & 0 \\
X_ {12} & 1 & 0 & 0 & 0 & 0 & 0 & 1 & 0 & 0 & 0 & 0 & 1 & 0 \\
Y_ {12} & 0 & 1 & 0 & 0 & 0 & 0 & 1 & 0 & 0 & 0 & 0 & 1 & 0 \\
Z_ {12} &0 & 0 & 1 & 0 & 0 & 0 & 1 & 0 & 0 & 0 & 0 & 1 & 0 \\
X_ {51} & 1 & 0 & 0 & 0 & 0 & 0 & 0 & 1 & 0 & 0 & 0 & 0 & 1 \\
Y_ {51} & 0 & 1 & 0 & 0 & 0 & 0 & 0 & 1 & 0 & 0 & 0 & 0 & 1 \\
Z_ {51} & 0 & 0 & 1 & 0 & 0 & 0 & 0 & 1 & 0 & 0 & 0 & 0 & 1 \\
X_ {63} & 1 & 0 & 0 & 0 & 0 & 0 & 0 & 0 & 1 & 0 & 0 & 0 & 1 \\
Y_ {63} & 0 & 1 & 0 & 0 & 0 & 0 & 0 & 0 & 1 & 0 & 0 & 0 & 1 \\
Z_ {63} & 0 & 0 & 1 & 0 & 0 & 0 & 0 & 0 & 1 & 0 & 0 & 0 & 1 \\
Q_ {26} & 0 & 0 & 0 & 0 & 1 & 1 & 0 & 0 & 0 & 1 & 0 & 0 & 0 \\
Q_ {14} & 0 & 0 & 0 & 0 & 1 & 0 & 1 & 0 & 0 & 1 & 0 & 0 & 0 \\
Q_ {31} & 0 & 0 & 0 & 0 & 1 & 0 & 0 & 1 & 0 & 0 & 1 & 0 & 0 \\
Q_ {65} & 0 & 0 & 0 & 0 & 1 & 0 & 0 & 0 & 1 & 0 & 1 & 0 & 0 \\
\end{array}
\right) 
$}
~,~
\eea
where the columns of the $P$-matrix correspond to the GLSM fields of the $Y^{2,4}(\mathbb{CP}^2)$ model.
$p_1, \dots p_{11}$ correspond to regular GLSM fields, whereas $o_1,o_2$ correspond to extra GLSM fields for the master space ${}^{\text{Irr}}\mathcal{F}^\flat$ of the $Y^{2,4}(\mathbb{CP}^2)$ model.
We will see later on how we have identified the extra GLSM fields $o_1,o_2$ from the toric diagram of the master space ${}^{\text{Irr}}\mathcal{F}^\flat$.

In order to obtain the toric diagram for the master space ${}^{\text{Irr}}\mathcal{F}^\flat$, we first summarize the $U(1)$ charges on the GLSM fields due to the $J$- and $E$-terms of the $Y^{2,4}(\mathbb{CP}^2)$ model.
These charges are summarized in the following $Q_{JE}$-matrix,
\beal{es13a03}
Q_{JE}=
\resizebox{0.475\textwidth}{!}{$
\left(
\begin{array}{ccc|cc|cc|cc|cc|cc}
p_1 & p_2 & p_3 & p_4 & p_5 & p_6 & p_7 & p_8 & p_9 & p_{10} & p_{11}  & o_1 & o_2\\
\hline
 0 & 0 & 0 & 0 & 1 & -1 & -1 & 0 & 0 & 0 & -1 & 1 & 0 \\
 1 & 1 & 1 & 0 & 1 & -1 & -1 & -1 & -1 & 0 & 0 & 0 & 0 \\
 0 & 0 & 0 & 1 & 1 & 0 & 0 & 0 & 0 & -1 & -1 & 0 & 0 \\
 0 & 0 & 0 & 0 & 1 & 0 & 0 & -1 & -1 & -1 & 0 & 0 & 1 \\
\end{array}
\right)
$}
~,~
\eea
where we note that all GLSM fields, including both extra GLSM fields, carry charges charges under the $J$- and $E$-terms.
In addition to the charges due to the $J$- and $E$-terms, the GLSM fields also carry charges due to the $D$-terms of the $Y^{2,4}(\mathbb{CP}^2)$ model. These are summarized in,
\beal{es13a03b}
Q_D=
\resizebox{0.475\textwidth}{!}{$
\left(
\begin{array}{ccc|cc|cc|cc|cc|cc}
p_1 & p_2 & p_3 & p_4 & p_5 & p_6 & p_7 & p_8 & p_9 & p_{10} & p_{11}  & o_1 & o_2\\
\hline
 0 & 0 & 0 & 0 & 0 & 0 & -1 & 0 & 0 & 1 & 0 & 0 & 0 \\
 0 & 0 & 0 & 0 & 0 & 0 & 0 & 0 & -1 & 0 & 1 & 0 & 0 \\
 0 & 0 & 0 & 0 & 0 & 1 & 0 & 0 & 0 & -1 & 0 & 0 & 0 \\
 0 & 0 & 0 & 0 & 0 & 0 & 0 & 1 & 0 & 0 & -1 & 0 & 0 \\
 0 & 0 & 0 & 0 & 0 & -1 & 0 & 0 & 1 & 0 & 0 & 0 & 0 \\
\end{array}
\right)
$}
~.~
\eea
We observe that the $Q_{JE}$- and $Q_{D}$-matrices together indicate that the mesonic moduli space $\mathcal{M}_{mes}$ has a symmetry of the form,
\beal{es13a04a}
SU(3)_{(x_1,x_2)} \times U(1)_{f} \times U(1)_R
~.~
\eea
The $SU(3)$ enhancement is due to the fact that the GLSM fields $(p_1,p_2,p_3)$ carry the same charges under the $J$-, $E$- and $D$-terms of the $Y^{2,4}(\mathbb{CP}^2)$ model.
In comparison, when we focus just on the master space ${}^{\text{Irr}}\mathcal{F}^\flat$, 
the $Q_{JE}$-matrix indicates that the global symmetry of ${}^{\text{Irr}}\mathcal{F}^\flat$ is 
\beal{es13a04}
SU(3)_{(x_1,x_2)} \times SU(2)_y \times SU(2)_z \times U(1)_{b_1} \times U(1)_{b_2} \times U(1)_{b_3} \times U(1)_{b_4} \times U(1)_R ~,~
\nn\\
\eea
where the total rank of the global symmetry of the master space ${}^{\text{Irr}}\mathcal{F}^\flat$ is as expected $G+3=9$. 
Here, we note that the additional $SU(2)$ enhancements are due to the fact that the pairs of GLSM fields $(p_6,p_7)$ and $(p_8,p_9)$ carry the same charges in the $Q_{JE}$-matrix.

%--------------------------------
\begin{table}[ht!]
\centering
\resizebox{\textwidth}{!}{
\begin{tabular}{|c|c|c|c|c|c|c|c|c|c|}
\hline
\; & $SU(3)_{[x_1,x_2]}$  & $SU(2)_{y}$ & $SU(2)_{z}$ & $U(1)_{b_1}$ & $U(1)_{b_2}$ & $U(1)_{b_3}$ & $U(1)_{b_4}$  & $U(1)_R$ & $fugacity$\\
\hline\hline
$p_1$ & $(+1,~~0)$   & $0$  & $0$  & $0$   & $0$      & $0$ & $0$     & $r_1$ &   $t_1=x_1 t$  \\
$p_2$ & $(-1,+1)$      & $0$  & $0$  & $0$   & $0$      & $0$ & $0$     & $r_2$ &   $t_2=x_1^{-1} x_2 t$  \\
$p_3$ & $(0,~-1)$      & $0$  & $0$  & $0$   & $0$      & $0$ & $0$     & $r_3$ &   $t_3=x_2^{-1} t$  \\
\hline
$p_4$ & $0$ & $0$& $0$   & $+1$  & $+1$      & $0$ & $0$     & $r_4$ &   $t_4=b_1 b_2 t$  \\
$p_5$ & $0$ & $0$ & $0$  & $-1$   & $+1$      & $0$ & $0$     & $r_5$ &   $t_5=b_1^{-1} b_2 t $  \\
\hline
$p_6$ & $0$ & $+1$ & $0$   & $0$ & $-1$  & $+1$ & $0$   & $r_6$ &    $t_6=y b_2^{-1} b_3 t$  \\
$p_7$ & $0$ & $-1$ & $0$    & $0$ & $-1$   & $+1$ & $0$  &  $r_7$ &   $t_7=y^{-1} b_2^{-1} b_3 t$  \\
\hline
$p_8$ & $0$  & $0$& $+1$     & $0$ & $0$   & $-1$ & $0$ & $r_8$ &     $t_8=z b_3^{-1} t$  \\
$p_9$ & $0$  & $0$& $-1$     & $0$ & $0$    & $-1$ & $0$ & $r_9$ &     $t_9=z^{-1} b_3^{-1} t$  \\
\hline
$p_{10}$ & $0$ & $0$ & $0$     & $0$ & $0$   & $0$ & $+1$ & $r_{10}$ &  $t_{10}= b_4 t$  \\
$p_{11}$ & $0$ & $0$ & $0$     & $0$ & $0$   & $0$ & $-1$ & $r_{11}$ &   $t_{11}= b_4^{-1} t$  \\
\hline
$o_1$ & $0$ & $0$ & $0$     & $0$ & $0$   & $0$ & $0$ & $0$ &   $u_1=1$  \\
$o_2$ & $0$ & $0$ & $0$     & $0$ & $0$   & $0$ & $0$ & $0$ &   $u_2=1$  \\
\hline
\end{tabular}
}
\caption{The global symmetry charges of the master space ${}^{\text{Irr}}\mathcal{F}^\flat$ on the GLSM fields $p_\alpha$ for the $Y^{2,4}(\mathbb{CP}^2)$.
\label{tex07}}
\end{table}
%--------------------------------

Given that the $J$- and $E$-terms in \eref{es13a01} are all binomial, the master space ${}^{\text{Irr}}\mathcal{F}^\flat$ is toric and has a $8$-dimensional toric diagram which is given by
\beal{es13a05}
G_t^{{}^{\text{Irr}}\mathcal{F}^\flat}=
\resizebox{0.475\textwidth}{!}{$
\left(
\begin{array}{ccc|cc|cc|cc|cc|cc}
\;  p_1 & p_2 & p_3 & p_4 & p_5 & p_6 & p_7 & p_8 & p_9 & p_{10} & p_{11}  & o_1 & o_2\\
\hline
 1 & 0 & 0 & 0 & 0 & 0 & 0 & 0 & 1 & 0 & 0 & 0 & 1 \\
 0 & 1 & 0 & 0 & 0 & 0 & 0 & 0 & 1 & 0 & 0 & 0 & 1 \\
 0 & 0 & 1 & 0 & 0 & 0 & 0 & 0 & 1 & 0 & 0 & 0 & 1 \\
 0 & 0 & 0 & 1 & 0 & 0 & 0 & 0 & 0 & 0 & 1 & 1 & 0 \\
 0 & 0 & 0 & 0 & 1 & 0 & 0 & 0 & 1 & 0 & 1 & 0 & 0 \\
 0 & 0 & 0 & 0 & 0 & 1 & 0 & 0 & -1 & 0 & 0 & 1 & -1 \\
 0 & 0 & 0 & 0 & 0 & 0 & 1 & 0 & -1 & 0 & 0 & 1 & -1 \\
 0 & 0 & 0 & 0 & 0 & 0 & 0 & 1 & -1 & 0 & 0 & 0 & 0 \\
 \hline
 1 & 1 & 1 & 1 & 1 & 1 & 1 & 1 & 1  & 1 & 1 & 2 & 2 \\
\end{array}
\right)
$}
~.~
\eea
Here, we note that the toric diagram is a convex polytope on a 8-dimensional hyperplane consisting of vertices corresponding to the GLSM fields $p_1, \dots, p_{11}$.
Two vertices are outside the 8-dimensional hyperplane and correspond to the extra GLSM fields $o_1$ and $o_2$.
We can identify these two extra GLSM fields $o_1$ and $o_2$ as an over-parameterization of the master space ${}^{\text{Irr}}\mathcal{F}^\flat$ for the $Y^{2,4}(\mathbb{CP}^2)$ model.
This is because, as we can see in \tref{tex07}, they do not carry any global symmetry charges and their absence from the parameterization of the master space ${}^{\text{Irr}}\mathcal{F}^\flat$ does not affect the algebraic description of ${}^{\text{Irr}}\mathcal{F}^\flat$ in terms of generators and defining first order relations.

We can observe this by calculating the Hilbert series using the symplectic quotient description of the master space ${}^{\text{Irr}}\mathcal{F}^\flat$.
The Hilbert series takes the following rational form,
\beal{es13a20}
&&
g(
t_\alpha, u_1, u_2;
{}^{\text{Irr}}\mathcal{F}^\flat
)
=
\frac{
P(t_\alpha,u_1,u_2)
}{
(1-u_1 t_1 t_6) (1-u_1 t_2 t_6) (1-u_1 t_3 t_6) (1-u_1 t_1 t_7) (1-u_1 t_2 t_7) 
}
\nn\\
&&
\hspace{1cm}
\times \frac{1}{
(1-u_1 t_3 t_7) (1-u_2 t_1 t_8) (1-u_2 t_2 t_8) (1-u_2 t_3 t_8) (1-u_2 t_1 t_9) (1-u_2 t_2 t_9)
}
\nn\\
&&
\hspace{1cm}
\times \frac{1}{
 (1-u_2 t_3 t_9) (1-u_2 t_4 t_{10}) (1-t_5 t_6 t_{10}) (1-t_5 t_7 t_{10}) (1-u_1 t_4 t_{11}) (1-t_5 t_8 t_{11})
}
\nn\\
&&
\hspace{1cm}
\times \frac{1}{
 (1-t_5 t_9 t_{11})
}
~,~
\eea
where the numerator $P(t_\alpha, u_1,u_2)$ contains $1886$ terms. 
The fugacities $t_\alpha$ count degrees in the regular GLSM fields $p_\alpha$, whereas fugacities $u_1$ and $u_2$ count degrees in the extra GLSM fields $o_1$ and $o_2$, respectively.
The corresponding plethystic logarithm takes the form,
\beal{es13a22}
&&
\text{PL}[
g(
t_\alpha,
u_1,u_2;
{}^{\text{Irr}}\mathcal{F}^\flat
)]
=
u_1 t_1 t_6+u_1 t_2 t_6+u_1 t_3 t_6+u_1 t_1 t_7+u_1 t_2 t_7 +u_1 t_3 t_7
\nn\\
&&
\hspace{1cm}
+u_2 t_1 t_8
+u_2 t_2 t_8 +u_2 t_3 t_8+u_2 t_1 t_9+u_2 t_2 t_9+u_2 t_3 t_9+u_2 t_4 t_{10}+u_1 t_4 t_{11}
\nn\\
&&
\hspace{1cm}
+t_5 t_{10} t_6
+t_5 t_7 t_{10}+t_5 t_8 t_{11}
+t_5 t_9 t_{11}-(u_1^2 t_1 t_2 t_6 t_7+u_1^2 t_1 t_3 t_6 t_7+u_1^2 t_2 t_3 t_6 t_7 
\nn\\
&&
\hspace{1cm}
+u_1 u_2 t_1 t_2 t_6 t_8
+u_1 u_2 t_1 t_3 t_6 t_8
+u_1 u_2 t_2 t_3 t_6 t_8
+u_1 u_2 t_1 t_2 t_7 t_8
+u_1 u_2 t_1 t_3 t_7 t_8 
\nn\\
&&
\hspace{1cm}
+u_1 u_2 t_2 t_3 t_7 t_8
+u_1 u_2 t_1 t_2 t_6 t_9
+u_1 u_2 t_1 t_3 t_6 t_9
+u_1 u_2 t_2 t_3 t_6 t_9
+u_1 u_2 t_1 t_2 t_7 t_9 
\nn\\
&&
\hspace{1cm}
+u_1 u_2 t_1 t_3 t_7 t_9
+u_1 u_2 t_2 t_3 t_7 t_9
+u_1 t_1 t_5 t_6 t_7 t_{10}+u_1 t_2 t_5 t_6 t_7 t_{10}+u_1 t_3 t_5 t_6 t_7 t_{10}
\nn\\
&&
\hspace{1cm}
+u_2^2 t_1 t_2 t_8 t_9
+u_2^2 t_1 t_3 t_8 t_9+u_2^2 t_2 t_3 t_8 t_9
+u_2 t_1 t_5 t_8 t_9 t_{11}+u_2 t_2 t_5 t_8 t_9 t_{11} 
\nn\\
&&
\hspace{1cm}
+u_2 t_3 t_5 t_8 t_9 t_{11})
+\dots
~.~
\eea
We note that the master space ${}^{\text{Irr}}\mathcal{F}^\flat$ has 18 generators that satisfy 24 first order relations.
Given that the plethystic logarithm is not finite, the master space is a non-complete intersection.
Furthermore, we note that if we set the fugacities corresponding to the extra GLSM fields to $u_1=1$ and $u_2=1$, the 18 generators still satisfy the same 24 first order relations according to the plethystic logarithm. 
We will see this in more detail when we explicitly construct the generators and first order relations later in this section.

First, let us illustrate that the Hilbert series calculated from the symplectic quotient description of the master space ${}^{\text{Irr}}\mathcal{F}^\flat$ is identical to the Hilbert series obtained from the quotienting ideal given by the $J$- and $E$-terms of the $Y^{2,4}(\mathbb{CP}^2)$ model.
We can express the chiral fields of the $Y^{2,4}(\mathbb{CP}^2)$ brane brick model as products of GLSM fields as follows, 
\beal{es13a30}
&
P_{35} = p_4 p_{11} o_1 ~,~
P_{24} = p_4 p_{10} o_2 ~,~
&
\nn\\
&
X_{46} = p_1 p_6 o_1 ~,~
Y_{46} = p_2 p_6 o_1 ~,~
Z_{46} = p_3 p_6 o_1 ~,~
&
\nn\\
&
X_{12} = p_1 p_7 o_1 ~,~
Y_{12} = p_2 p_7 o_1 ~,~
Z_{12} = p_3 p_7 o_1 ~,~
&
\nn\\
&
X_{51} = p_1 p_8 o_2 ~,~
Y_{51} = p_2 p_8 o_2 ~,~
Z_{51} = p_3 p_8 o_2 ~,~
&
\nn\\
&
X_{63} = p_1 p_9 o_2 ~,~
Y_{63} = p_2 p_9 o_2 ~,~
Z_{63} = p_3 p_9 o_2 ~,~
&
\nn\\
&
Q_{26} = p_5 p_6 p_{10} ~,~
Q_{14} = p_5 p_7 p_{10} ~,~
Q_{31} = p_5 p_8 p_{11} ~,~
Q_{65} = p_5 p_9 p_{11} 
~.~
\eea
Under primary decomposition of the $J$- and $E$-terms in \eref{es13a01}, the coherent component takes the following form,
\beal{es13a35}
&&
\mathcal{I}_{JE}^{\text{Irr}} =
\big\langle
~
\nn\\&&
\hspace{1cm}
Q_{65} Y_{51} - Y_{63} Q_{31} ~,~
X_{12} Q_{26} -Q_{14} X_{46}~,~
Q_{65} X_{51}- X_{63} Q_{31} ~,~
Y_{12} Q_{26}-Q_{14} Y_{46}~,~
\nn\\&&
\hspace{1cm}
Z_{12} Q_{26}-Q_{14} Z_{46}~,~
Z_{63} Q_{31}-Q_{65} Z_{51}~,~
X_{63} Y_{51}-X_{51} Y_{63}~,~
Z_{63} Y_{51}-Z_{51} Y_{63}~,~
\nn\\&&
\hspace{1cm}
Y_{46} Z_{51}-Z_{46} Y_{51}~,~
Y_{12} Z_{51}-Z_{12} Y_{51}~,~
X_{12} Y_{46}-Y_{12} X_{46}~,~
X_{63} Y_{46}-X_{46} Y_{63}~,~
\nn\\&&
\hspace{1cm}
X_{51} Y_{46}-X_{46} Y_{51}~,~
Z_{63} Y_{46}-Z_{46} Y_{63}~,~
Z_{46} X_{12}-Z_{12} X_{46}~,~
Z_{46} X_{63}-Z_{63} X_{46}~,~
\nn\\&&
\hspace{1cm}
Y_{12} X_{63}-X_{12} Y_{63}~,~
Z_{12} X_{63}-Z_{63} X_{12}~,~
Z_{46} X_{51}-Z_{51} X_{46}~,~
Y_{12} X_{51}-X_{12} Y_{51}~,~
\nn\\&&
\hspace{1cm}
Z_{63} X_{51}-X_{63} Z_{51}~,~
Z_{12} X_{51}-X_{12} Z_{51}~,~
Y_{12} Z_{46}-Z_{12} Y_{46}~,~
Z_{63} Y_{12}-Z_{12} Y_{63}~,~
\nn\\&&
\hspace{1cm}
P_{24} Q_{65} Y_{46}-P_{35} Q_{26} Y_{63}~,~
P_{24} Q_{31} Y_{46}-P_{35} Q_{26} Y_{51}~,~
P_{35} Q_{26} X_{63}-P_{24} Q_{65} X_{46}~,~
\nn\\&&
\hspace{1cm}
P_{35} Q_{14} X_{63}-P_{24} Q_{65} X_{12}~,~
P_{35} Q_{26} X_{51}-P_{24} Q_{31} X_{46}~,~
P_{35} Q_{14} X_{51}-P_{24} Q_{31} X_{12}~,~
\nn\\&&
\hspace{1cm}
P_{24} Q_{31} Z_{46}-P_{35} Q_{26} Z_{51}~,~
P_{35} Z_{63} Q_{26}-P_{24} Q_{65} Z_{46}~,~
P_{24} Q_{65} Y_{12}-P_{35} Q_{14} Y_{63}~,~
\nn\\&&
\hspace{1cm}
P_{24} Q_{31} Y_{12}-P_{35} Q_{14} Y_{51}~,~
Z_{12} P_{24} Q_{31}-P_{35} Q_{14} Z_{51}~,~
P_{35} Z_{63} Q_{14}-Z_{12} P_{24} Q_{65}
~
\big\rangle
~.~
\nn\\
\eea
We note that the coherent component is still made of binomial relations. 
Accordingly, the master space ${}^{\text{Irr}}\mathcal{F}^\flat$ is given by the following toric variety
\beal{es13a36}
{}^{\text{Irr}}\mathcal{F}^\flat
= 
\text{Spec}~
\mathbb{C}^{18}[
P_{ij},
Q_{ij},
X_{ij},
Y_{ij},
Z_{ij}
]
/ \mathcal{I}_{JE}^{\text{Irr}}
~,~
\eea
whose corresponding Hilbert series can be calculated using \textit{Macaulay2} \cite{M2}. 
By an appropriate map between fugacities corresponding to chiral fields and fugacities corresponding to GLSM fields, following the relations in \eref{es13a35}, one can show that the two Hilbert series are identical.

Given that the global symmetry of the master space ${}^{\text{Irr}}\mathcal{F}^\flat$ is $SU(3)_{(x_1,x_2)} \times SU(2)_y \times SU(2)_z \times U(1)_{b_1} \times U(1)_{b_2} \times U(1)_{b_3} \times U(1)_{b_4} \times U(1)_R$, we can introduce a fugacity map that maps global symmetry fugacities $(x_1,x_2,y,b_1,b_2,t)$ to fugacities corresponding to GLSM fields $(t_\alpha,u_1,u_2)$.
This map is given by,
\beal{es13a40}
&
t = t_1^{1/11} t_2^{1/11} t_3^{1/11} t_4^{1/11} t_5^{1/11} t_6^{1/11} t_7^{1/11} t_8^{1/11} t_9^{1/11} t_{10}^{1/11} t_{11}^{1/11} u_1^{2/11} u_2^{2/11}
~,~
&
\nn\\
&
x_1 = \frac{t_1^{2/3}}{t_2^{1/3} t_3^{1/3}}~,~ 
x_2 = \frac{t_1^{1/3} t_2^{1/3}}{t_3^{2/3}}~,~ 
y = \frac{t_6^{1/2}}{t_7^{1/2}}~,~
z = \frac{t_8^{1/2}}{t_9^{1/2}}~,~
&
\nn\\
&
b_1=\frac{u_1^{9/22} u_2^{9/22} t_1^{4/33} t_2^{4/33} t_3^{4/33} t_4^{5/11}}{t_5^{6/11} t_6^{1/22} t_7^{1/22} t_8^{1/22} t_9^{1/22} t_{10}^{1/22} t_{11}^{1/22}}~,~
b_2=\frac{t_4^{4/11} t_5^{4/11} t_{10}^{4/11} t_{11}^{4/11} }{u_1^{3/11} u_2^{3/11} t_1^{10/33} t_2^{10/33} t_3^{10/33} t_6^{3/22} t_7^{3/22} t_8^{3/22} t_9^{3/22}}~,~
&
\nn\\
&
b_3=\frac{u_1^{4/11} t_4^{2/11} t_5^{2/11} t_6^{2/11} t_7^{2/11} t_{10}^{2/11} t_{11}^{2/11}}{u_2^{7/11} t_1^{5/33} t_2^{5/33} t_3^{5/33} t_8^{7/22} t_9^{7/2}}~,~
b_4=\frac{u_2^{1/2} t_{10}^{1/2}}{u_1^{1/2} t_{11}^{1/2}}
~.~
\eea
In terms of the above fugacity map, we can rewrite the refined Hilbert series in \eref{es13a20} in terms of characters of irreducible representations of the global symmetry of the master space ${}^{\text{Irr}}\mathcal{F}^\flat$.
Note that even if we set the fugacities for the extra GLSM fields to $u_1=1$ and $u_2=1$, the corresponding fugacity map in \eref{es13a40} will allow us to rewrite the refined Hilbert series in \eref{es13a20} in terms of the same characters of irreducible representations of the global symmetry of the master space ${}^{\text{Irr}}\mathcal{F}^\flat$.

The corresponding highest weight generating function \cite{Hanany:2014dia} is given by
\beal{es13a42}
&&
h(t,\mu_i,\nu,\kappa,b_j;{}^{\text{Irr}}\mathcal{F}^\flat)
=
\frac{
1-\mu_1 \nu \kappa b_2 t^7
}{
(1-b_1 b_2 b_4^{-1} t^2) (1-b_1 b_2 b_4 t^2) (1-\mu_1 \nu b_2^{-1} b_3 t^2) 
}
\nn\\
&&
\hspace{1cm}
\times \frac{1}{
(1-\mu_1 \kappa  b_3^{-1} t^2) (1-\nu b_1^{-1} b_3 b_4 t^3) (1-\kappa b_1^{-1} b_2 b_3^{-1} b_4^{-1} t^3)
}
~,~
\eea
where $ \mu_1^{m_1} \mu_2^{m_2}  \nu^{n} \kappa^{k}  \sim [m_1,m_2]_{SU(3)_{(x_1,x_2)}} [n]_{SU(2)_{y}} [k]_{SU(2)_{z}}$.
The plethystic logarithm of the refined Hilbert series in terms of global symmetry fugacities takes the form,
\beal{es13a45}
&&
PL[g (t,x_i,y,b_j;{}^{\text{Irr}}\mathcal{F}^\flat )]
=
[0,0;0;0] b_1 b_2 b_4^{-1} t^2+[0,0;0;0] b_1 b_2 b_4 t^2+[1,0;1;0]b_2^{-1} b_3 t^2
\nn\\
&&
\hspace{1cm}
+[1,0;0;1]b_3^{-1} t^2+[0,0;1;0]b_1^{-1} b_3 b_4 t^3+[0,0;0;1] b_1^{-1} b_2 b_3^{-1} b_4^{-1} t^3
\nn\\
&&
\hspace{1cm}
- ([0,1;0;0] b_2^{-2} b_3^2 t^4
+[0,1;0;0]b_3^{-2} t^4 +[0,1;1;1] b_2^{-1} t^4
+[1,0;0;0] b_1^{-1} b_2^{-1} b_3^2 b_4 t^5
\nn\\
&&
\hspace{1cm}
+[1,0;0;0]b_1^{-1} b_2 b_3^{-2} b_4^{-1} t^5)
+
\dots
~.~
\eea
From the plethystic logarithm, we identify the generators of the master space ${}^{\text{Irr}}\mathcal{F}^\flat$ 
and the associated irreducible representations under the global symmetry of the master space ${}^{\text{Irr}}\mathcal{F}^\flat$ as follows,
\beal{es13a46}
A = P_{35}  &~~\leftrightarrow~~& + [0,0;0;0]  b_1 b_2 b_4^{-1} t^2
~,~
\\
B =P_{24} &~~\leftrightarrow~~&   +[0,0;0;0] b_1 b_2 b_4 t^2
~,~
\\
C_{ij} = \left(
\ba{ccc}
X_{46} & Y_{46} & Z_{46} \\
X_{12} & Y_{12} & Z_{12} \\ 
\ea
\right)
&~~\leftrightarrow~~&
+[1,0;1;0]b_2^{-1} b_3 t^2
~.~
\\
D_{ij} = \left(
\ba{ccc}
X_{51} & Y_{51} & Z_{51} \\
X_{63} & Y_{63} & Z_{63} \\ 
\ea
\right)
&~~\leftrightarrow~~&
+[1,0;0;1]b_3^{-1} t^4
~.~
\\
E_{i}=
\left(
\ba{cc}
Q_{26} & Q_{14}  \\
\ea
\right)
&~~\leftrightarrow~~&
+[0,0;1;0]b_1^{-1} b_3 b_4 t^3
\\
F_{i}=
\left(
\ba{cc}
Q_{31} & Q_{65}  \\
\ea
\right)
&~~\leftrightarrow~~&
+[0,0;0;1] b_1^{-1} b_2 b_3^{-1} b_4^{-1} t^3
\eea
The first order relations formed by the generators are given by,
\beal{es13a47}
M^{m}=\frac{1}{2} \epsilon^{mkl} \epsilon^{ij}C_{i k} C_{jl}=0 &~~\leftrightarrow~~& -[0,1;0;0]b_2^{-2} b_3^{2} t^4
~,~
\\
N^{m}=\frac{1}{2} \epsilon^{mkl} \epsilon^{ij}D_{i k} D_{jl}=0 &~~\leftrightarrow~~& -[0,1;0;0]b_3^{-2} t^4
~,~
\\
P_{ik}^{m}=\frac{1}{2} \epsilon^{mjl} C_{i j} D_{kl}=0 &~~\leftrightarrow~~& -[0,1;1;1] b_2^{-1} t^4
~,~
\\
Q_{k}=\epsilon^{ij} C_{ik} E_{j}=0 &~~\leftrightarrow~~& -[1,0;0;0] b_1^{-1} b_2^{-1} b_3^2 b_4 t^5
~,~
\\
R_{k}=\epsilon^{ij} D_{ik} F_{j}=0 &~~\leftrightarrow~~& -[1,0;0;0]b_1^{-1} b_2 b_3^{-2} b_4^{-1} t^5
~.~
\eea
\tref{tex08} summarizes the generators of the master space ${}^{\text{Irr}}\mathcal{F}^\flat$ and shows them in terms of GLSM fields and their global symmetry charges. 

%--------------------------------
\begin{table}[ht!]
\centering
\resizebox{\textwidth}{!}{
\begin{tabular}{|c|c|c|c|c|c|c|c|c|c|c|}
\hline
generators  & GLSM fields & $SU(3)_{(x_1,x_2)}$ & $SU(2)_{y}$ & $SU(2)_{z}$ & $U(1)_{b_1}$& $U(1)_{b_2}$& $U(1)_{b_3}$& $U(1)_{b_4}$ & $U(1)_R$\\
\hline\hline
$A= P_{35}$ & $p_4 p_{11} o_1 $ & $0$ &$0$&$0$& $+1$ & $+1$& $0$& $-1$& $r_4+r_{11}$\\
\hline
$B=P_{24}$ & $p_4 p_{10} o_2 $ & $0$ &$0$&$0$& $+1$ & $+1$& $0$& $+1$& $r_4+r_{10}$\\
\hline
$C_{11}= X_{46}$ & $p_1 p_6 o_1 $ & $\left(+1,~~0\right)$ &$+1$&$0$&$0$ &$-1$&$+1$&$0$& $r_1+r_6$\\
$C_{12}= Y_{46}$ & $p_2 p_6 o_1 $ & $\left(-1,+1\right)$ &$+1$&$0$&$0$ &$-1$&$+1$&$0$& $r_2+r_6$\\
$C_{13}= Z_{46}$ & $p_3 p_6 o_1 $ & $\left(0,~-1\right)$ &$+1$&$0$&$0$ &$-1$&$+1$&$0$& $r_3+r_6$\\
$C_{21}= X_{12}$ & $p_1 p_7 o_1 $ & $\left(+1,~~0\right)$ &$-1$&$0$&$0$ &$-1$&$+1$&$0$& $r_1+r_7$\\
$C_{22}= Y_{12}$ & $p_2 p_7 o_1 $ & $\left(-1,+1\right)$ &$-1$&$0$&$0$ &$-1$&$+1$&$0$& $r_2+r_7$\\
$C_{23}= Z_{12}$ & $p_3 p_7 o_1 $ & $\left(0,~-1\right)$ &$-1$&$0$&$0$ &$-1$&$+1$&$0$& $r_3+r_7$\\
\hline
$D_{11}= X_{51}$ & $p_1 p_8 o_2 $ & $\left(+1,~~0\right)$ &$0$&$+1$&$0$ &$0$&$-1$&$0$& $r_1+r_8$\\
$D_{12}= Y_{51}$ & $p_2 p_8 o_2 $ & $\left(-1,+1\right)$ &$0$&$+1$&$0$ &$0$&$-1$&$0$& $r_2+r_8$\\
$D_{13}= Z_{51}$ & $p_3 p_8 o_2 $ & $\left(0,~-1\right)$ &$0$&$+1$&$0$ &$0$&$-1$&$0$& $r_3+r_8$\\
$D_{21}= X_{63}$ & $p_1 p_9 o_2 $ & $\left(+1,~~0\right)$ &$0$&$-1$&$0$ &$0$&$-1$&$0$& $r_1+r_9$\\
$D_{22}= Y_{63}$ & $p_2 p_9 o_2 $ & $\left(-1,+1\right)$ &$0$&$-1$&$0$ &$0$&$-1$&$0$& $r_2+r_9$\\
$D_{23}= Z_{63}$ & $p_3 p_9 o_2 $ & $\left(0,~-1\right)$ &$0$&$-1$&$0$ &$0$&$-1$&$0$& $r_3+r_9$\\
\hline
$E_{1}= Q_{26}$ & $p_5 p_6 p_{10} $  &$0$&$-1$&$0$ &$-1$ &$0$ &$+1$ &$+1$ & $r_5+r_6$\\
$E_{2}= Q_{14}$ & $p_5 p_7 p_{10} $  &$0$&$-1$&$0$ &$-1$ &$0$ &$+1$ &$+1$ & $r_5+r_7$\\
\hline
$F_{1}= Q_{31}$ & $p_5 p_8 p_{11} $  &$0$&$0$&$0$&$-1$&$+1$&$-1$&$-1$ & $r_5+r_8$\\
$F_{2}= Q_{65}$ & $p_5 p_9 p_{11} $  &$0$&$0$&$0$&$-1$&$+1$&$-1$&$-1$ & $r_5+r_9$\\
\hline
\end{tabular}}
\caption{
Generators of the master space ${}^{\text{Irr}}\mathcal{F}^\flat$ for the  $Y^{2,4}(\mathbb{CP}^2)$ model with the global symmetry charges. 
\label{tex08}
}
\end{table}
%--------------------------------

We have identified the master space ${}^{\text{Irr}}\mathcal{F}^\flat$ of the $Y^{2,4}(\mathbb{CP}^2)$ model as a 9-dimensional toric variety. 
By calculating the plethystic logarithm of the Hilbert series in \eref{es13a45}, we further identified the master space ${}^{\text{Irr}}\mathcal{F}^\flat$ as a non-complete intersection. 
However, when we unrefine the Hilbert series by setting the fugacities $t_\alpha = t$, $u_1=1$ and $u_2=1$, we obtain
\beal{es13a21}
&&
g(
t;
{}^{\text{Irr}}\mathcal{F}^\flat
)
=
(1-18 t^4-6 t^5+52 t^6+12 t^7-60 t^8+52 t^9+42 t^{10}-226 t^{11}-82 t^{12}
\nn\\
&&
\hspace{1cm}
+345 t^{13}+158 t^{14}
-242 t^{15}-93 t^{16}+58 t^{17}-110 t^{18}-6 t^{19}+212 t^{20}+46 t^{21}
\nn\\
&&
\hspace{1cm}
-138 t^{22}-50 t^{23}+40 t^{24}+18 t^{25}
-4 t^{26}-t^{29})\times
\frac{1}{ \left(1-t^2\right)^{14} \left(1-t^3\right)^{4}} ~,~
\eea
where we can see that the unrefined Hilbert series in rational form has a numerator, which is not palindromic.
This indicates, under a theorem by Stanley \cite{stanley1978hilbert}, that the corresponding master space ${}^{\text{Irr}}\mathcal{F}^\flat$ of the $Y^{2,4}(\mathbb{CP}^2)$ model is \textit{not} Calabi-Yau. 
This is another case, where the master space ${}^{\text{Irr}}\mathcal{F}^\flat$ of a brane brick model is toric, but not Calabi-Yau, whereas the mesonic moduli space is a toric Calabi-Yau 4-fold.
\\

%=================================================================
\section{Observations and Discussions \label{sec:05}}

In this work, we have for the first time systematically studied the master space of brane brick models which realize a large class of $2d$ $(0,2)$ supersymmetric gauge theories corresponding to toric Calabi-Yau 4-folds.
The master space for brane brick models is defined as the space of chiral fields subject to the $J$- and $E$-terms constraints and the non-abelian part of the gauge symmetry.

In this work, we focused on the master spaces for brane brick models with $U(1)^G$ gauge groups.
The dimension of the master space is then $G+3$, where $4$ come from the mesonic and $G-1$ come from the baryonic directions of the master space. 
As for $4d$ $\mathcal{N}=1$ supersymmetric gauge theories realized by brane tilings corresponding to toric Calabi-Yau 3-folds \cite{Hanany:2005ve, Franco:2005rj, Hanany:2010zz, Forcella:2008bb, Forcella:2008eh, Forcella:2009bv, Zaffaroni:2008zz,Forcella:2008ng}, $U(1)^{G-1}$ decouple from the gauge symmetry and contribute as baryonic directions to the master space. 
The global symmetry of the brane brick model including contributions from these baryonic directions is a rank $G+3$ symmetry. Since the master space is larger than the mesonic moduli space, the mesonic flavor and $U(1)_R$ symmetry are contained in the full global symmetry of the master space. 
We saw in some of the examples studied in this work that the global symmetry of the master space can be further enhanced, containing non-abelian symmetry factors that have equal or even higher rank than non-abelian symmetry factors of the mesonic symmetry. 
Such enhancements of the global symmetry of the master space are not new, because similar enhancements of the global symmetry of the master space have been observed for brane tilings \cite{Hanany:2012vc}.

Even though master spaces of brane brick models have many similarities to master spaces of brane tilings, 
we have observed in our work that there are also significant differences. 
In the following, we summarize the two main new features that we have observed in this work for master spaces of abelian brane brick models:
\begin{itemize}

\item \textbf{Extra GLSM Fields.}
With $G$ $U(1)$ gauge groups, the master spaces for brane brick models are $(G+3)$-dimensional and toric.
The toric diagram for the master spaces is a convex polytope and consists of vertices that are on a $(G+2)$-dimensional hyperplane.
The vertices in the toric diagram correspond to GLSM fields of the brane brick model.
For certain brane brick models, some of the vertices in the toric diagram are located outside the $(G+2)$-dimensional hyperplane. 
We refer to the GLSM fields corresponding to these vertices as \textit{extra GLSM fields}. 
When we express the generators of the master space in terms of GLSM fields of the brane brick model, we can either include or exclude the extra GLSM fields. 
Either way, the set of defining first order relations amongst the generators remains unaffected indicating that the extra GLSM fields act as an over-parameterization of the master space. 
Such extra GLSM fields have appeared also in relation to the mesonic moduli spaces of abelian brane brick models \cite{Franco:2015tya,Franco:2015tna}.
We note however that the extra GLSM fields for the master space are not necessary related to the extra GLSM fields that appear for the mesonic moduli space. For some examples, we have observed that there are less extra GLSM fields for the master space than for the mesonic moduli space.

For example, for the brane brick model corresponding to $Q^{1,1,1}$ in section \sref{sec:043}, the mesonic moduli space exhibits two extra GLSM fields, while for the master space there is only one extra GLSM field. In fact, it appears that the vertex corresponding to one of the two extra GLSM fields in the mesonic moduli space becomes an additional extremal point of the toric diagram of the master space.

\item \textbf{Master Spaces that are not Calabi-Yau.}
The master space of brane brick models is toric and Calabi-Yau.
It is toric because the $J$- and $E$-terms of the brane brick model are binomial relations and under primary decomposition the coherent component remains a binomial ideal \cite{sturmfels1996grobner}.
We also identify the master space as Calabi-Yau by calculating the Hilbert series of the master space, which in rational form has a numerator that is palindromic. 
By Stanley's theorem \cite{stanley1978hilbert}, when the numerator of the Hilbert series in rational form is palindromic, the corresponding coordinate ring is Gorenstein and the variety is Calabi-Yau. 

When however the master space is over-parameterized by extra GLSM fields, we observe that the master space is not anymore Calabi-Yau.
The Hilbert series of the quotient ring describing the coherent component of the master space can be directly computed from the reduced binomial $J$- and $E$-terms using \textit{Macaulay2} \cite{M2}. 
We observe that when the master space has extra GLSM fields, its Hilbert series in rational form has a non-palindromic numerator.

The simplest example that exhibits this property is the master space for the brane brick model corresponding to $Q^{1,1,1}$.
The master space takes the form,
\beal{es50a01}
{}^{\text{Irr}}\mathcal{F}^\flat = \text{Spec} ~\mathbb{C}^{10}[X_{ij},Y_{ij}] / \mathcal{I}_{JE}^{\text{Irr}} ~,~
\eea
where the quotient is given by the reduced binomial $J$- and $E$-terms,
\beal{es50a02}
&&
\mathcal{I}_{JE}^{\text{Irr}} = 
\big\langle
~
X_{24} Y_{41}-Y_{23} X_{31}~,~
Y_{24} X_{41}- X_{23} Y_{31} ~,~
X_{24} X_{41}-X_{23} X_{31}~,~
\nn\\
&&
\hspace{1.75cm}
Y_{24} Y_{41}-Y_{23} Y_{31} ~,~
X_{23} Y_{41}- Y_{23} X_{41}~,~
X_{24} Y_{31}- Y_{24} X_{31}
~
\big\rangle
~,~
\eea
as originally shown in \eref{es12a35}. 
Note that the ideal in \eref{es50a02} defines the coherent component of the master space under primary decomposition of the original $J$- and $E$-terms.
When we grade all chiral fields $X_{ij},Y_{ij}$ the same and count their degrees with the same fugacity $\bar{t}$, the Hilbert series of the master space ${}^{\text{Irr}}\mathcal{F}^\flat$ obtained using \textit{Macaulay2} takes the form,
\beal{es50a03}
g(\bar{t}; {}^{\text{Irr}}\mathcal{F}^\flat) 
= 
\frac{
1 - 6 \bar{t}^2 + 8 \bar{t}^3 - 3 \bar{t}^4
}{
(1 - \bar{t})^8
}
~.~
\eea
We clearly see that the numerator of the Hilbert series is not palindromic.

We can also express the chiral fields $X_{ij},Y_{ij}$ in terms of GLSM fields, including the extra GLSM field $o$, as follows
\beal{es50a10}
&
X_{12} = p_1 ~,~
Y_{12} = p_2 ~,~
X_{31} = p_3 p_7 ~,~
Y_{31} = p_4 p_7 ~,~
X_{41} = p_5 p_7 ~,~
Y_{41} = p_6 p_7 ~,~
&
\nn\\
&
X_{24} = p_3 o ~,~
Y_{24} = p_4 o ~,~
X_{23} = p_5 o ~,~
Y_{23} = p_6 o ~.~
&
~,~
\eea
as first shown in \eref{es12a30}.
Using fugacities $t_\alpha$ for GLSM fields $p_\alpha$ and fugacity $u$ for extra GLSM field $o$, we can use the symplectic quotient description of the master space, 
\beal{es50a11}
{}^{\text{Irr}}\mathcal{F}^\flat = \mathbb{C}^c // Q_{JE}
~,~
\eea
with $c$ being the number of GLSM fields and 
\beal{es50a12}
Q_{JE}=
\resizebox{0.285\textwidth}{!}{$
\left(
\begin{array}{cc|cccc|c|c}
p_1 & p_2 & p_3 & p_4 & p_5 & p_6 & p_7 & o  \\
\hline
0 & 0 & -1 & -1 & -1 & -1 & 1 & 1 
\end{array}
\right)
$}
~,~
\eea
in order to calculate the following unrefined Hilbert series for the master space ${}^{\text{Irr}}\mathcal{F}^\flat$,
\beal{es50a13}
g(t_\alpha = t, u=t; {}^{\text{Irr}}\mathcal{F}^\flat) 
= 
\frac{
1 - 6 t^4 + 8 t^6 - 3 t^8
}{
(1-t)^2 (1-t^2)^8
}
~.~
\eea
We note that here again, the numerator of the Hilbert series is not palindromic.
Taken into account the interpretation that the extra GLSM fields correspond to an over-parameterization of the master space, we can set the fugacity for the extra GLSM field to $u=1$ to obtain,
\beal{es50a14}
g(t_\alpha = t, u=1; {}^{\text{Irr}}\mathcal{F}^\flat) 
= 
\frac{
1 - 6 t^3 + 4 t^4 + 3 t^5 - t^6 - t^7
}{
(1 - t)^6 (1 - t^2)^4
} ~.~
\eea
The numerator of the Hilbert series above is still not palindromic. 
Comparing the unrefined Hilbert series of the master space $ {}^{\text{Irr}}\mathcal{F}^\flat$ in \eref{es50a03}, \eref{es50a13} and \eref{es50a14}, we conclude that the master space for the abelian brane brick model corresponding to $Q^{1,1,1}$ is toric but \textit{not} Calabi-Yau.

In this work, we make similar observations for the master space for the abelian brane brick model corresponding to $Y^{2,4}(\mathbb{CP}^2)$.

\end{itemize}

The above features that we have observed for master spaces of brane brick models realizing $2d$ $(0,2)$ supersymmetric gauge theories have not been observed for master spaces of brane tilings realizing $4d$ $\mathcal{N}=1$ supersymmetric gauge theories \cite{Hanany:2005ve, Franco:2005rj, Hanany:2010zz, Forcella:2008bb, Forcella:2008eh, Forcella:2009bv, Zaffaroni:2008zz,Forcella:2008ng}.
We believe that with the increase of dimensionality of the probed toric Calabi-Yau singularity, the master spaces of the worldvolume theories living on the probe branes not only increase in their dimension, but also exhibit much richer and surprising algebro-geometric features. 
We plan to explore and to identify the origin of these features in future work. 
\\

%======================================================================
\section*{Acknowledgements}

R.-K. S. is supported by a Basic Research Grant of the National Research Foundation of Korea (NRF-2022R1F1A1073128).
He is also supported by a Start-up Research Grant for new faculty at UNIST (1.210139.01), a UNIST AI Incubator Grant (1.230038.01) and UNIST UBSI Grants (1.220123.01, 1.230065.01), as well as an Industry Research Project (2.220916.01) funded by Samsung SDS in Korea.  
He is also partly supported by the BK21 Program (``Next Generation Education Program for Mathematical Sciences'', 4299990414089) funded by the Ministry of Education in Korea and the National Research Foundation of Korea (NRF).
R.-K. S. is grateful to Per Berglund, Sebastian Franco, Dongwook Ghim, Amihay Hanany, Yang-Hui He and Sangmin Lee for discussions on related topics. 
He is also grateful to the Simons Center for Geometry and Physics at Stony Brook University for hospitality during the final stages of this work.
\\

%======================================================================
\bibliographystyle{JHEP}
\bibliography{mybib}

\providecommand{\href}[2]{#2}\begingroup\raggedright\begin{thebibliography}{10}

\bibitem{Franco:2015tna}
S.~Franco, D.~Ghim, S.~Lee, R.-K. Seong and D.~Yokoyama, \emph{{2d (0,2) Quiver
  Gauge Theories and D-Branes}},
  \href{http://dx.doi.org/10.1007/JHEP09(2015)072}{\emph{JHEP} {\bf 09} (2015)
  072}, [\href{http://arxiv.org/abs/1506.03818}{{\tt 1506.03818}}].

\bibitem{Franco:2015tya}
S.~Franco, S.~Lee and R.-K. Seong, \emph{{Brane Brick Models, Toric Calabi-Yau
  4-Folds and 2d (0,2) Quivers}},
  \href{http://dx.doi.org/10.1007/JHEP02(2016)047}{\emph{JHEP} {\bf 02} (2016)
  047}, [\href{http://arxiv.org/abs/1510.01744}{{\tt 1510.01744}}].

\bibitem{Franco:2016fxm}
S.~Franco, S.~Lee and R.-K. Seong, \emph{{Orbifold Reduction and 2d (0,2) Gauge
  Theories}}, \href{http://dx.doi.org/10.1007/JHEP03(2017)016}{\emph{JHEP} {\bf
  03} (2017) 016}, [\href{http://arxiv.org/abs/1609.07144}{{\tt 1609.07144}}].

\bibitem{Franco:2016nwv}
S.~Franco, S.~Lee and R.-K. Seong, \emph{{Brane brick models and 2d (0, 2)
  triality}}, \href{http://dx.doi.org/10.1007/JHEP05(2016)020}{\emph{JHEP} {\bf
  05} (2016) 020}, [\href{http://arxiv.org/abs/1602.01834}{{\tt 1602.01834}}].

\bibitem{Franco:2016qxh}
S.~Franco, S.~Lee, R.-K. Seong and C.~Vafa, \emph{{Brane Brick Models in the
  Mirror}}, \href{http://dx.doi.org/10.1007/JHEP02(2017)106}{\emph{JHEP} {\bf
  02} (2017) 106}, [\href{http://arxiv.org/abs/1609.01723}{{\tt 1609.01723}}].

\bibitem{Franco:2021ixh}
S.~Franco, A.~Mininno, A.~M. Uranga and X.~Yu, \emph{{2d $ \mathcal{N} $ = (0,
  1) gauge theories and Spin(7) orientifolds}},
  \href{http://dx.doi.org/10.1007/JHEP03(2022)150}{\emph{JHEP} {\bf 03} (2022)
  150}, [\href{http://arxiv.org/abs/2110.03696}{{\tt 2110.03696}}].

\bibitem{Franco:2018qsc}
S.~Franco and A.~Hasan, \emph{{$3d$ printing of $2d$ $
  \mathcal{N}=\left(0,2\right) $ gauge theories}},
  \href{http://dx.doi.org/10.1007/JHEP05(2018)082}{\emph{JHEP} {\bf 05} (2018)
  082}, [\href{http://arxiv.org/abs/1801.00799}{{\tt 1801.00799}}].

\bibitem{Franco:2022gvl}
S.~Franco and R.-K. Seong, \emph{{Fano 3-folds, reflexive polytopes and brane
  brick models}}, \href{http://dx.doi.org/10.1007/JHEP08(2022)008}{\emph{JHEP}
  {\bf 08} (2022) 008}, [\href{http://arxiv.org/abs/2203.15816}{{\tt
  2203.15816}}].

\bibitem{Franco:2022isw}
S.~Franco, D.~Ghim and R.-K. Seong, \emph{{Brane brick models for the
  Sasaki-Einstein 7-manifolds
  Y$^{p,k}$(\ensuremath{\mathbb{C}}\ensuremath{\mathbb{P}}$^{1}$\texttimes{}
  \ensuremath{\mathbb{C}}\ensuremath{\mathbb{P}}$^{1}$) and
  Y$^{p,k}$(\ensuremath{\mathbb{C}}\ensuremath{\mathbb{P}}$^{2}$)}},
  \href{http://dx.doi.org/10.1007/JHEP03(2023)050}{\emph{JHEP} {\bf 03} (2023)
  050}, [\href{http://arxiv.org/abs/2212.02523}{{\tt 2212.02523}}].

\bibitem{Hanany:2005ve}
A.~Hanany and K.~D. Kennaway, \emph{{Dimer models and toric diagrams}},
  \href{http://arxiv.org/abs/hep-th/0503149}{{\tt hep-th/0503149}}.

\bibitem{Franco:2005rj}
S.~Franco, A.~Hanany, K.~D. Kennaway, D.~Vegh and B.~Wecht, \emph{{Brane Dimers
  and Quiver Gauge Theories}},
  \href{http://dx.doi.org/10.1088/1126-6708/2006/01/096}{\emph{JHEP} {\bf 01}
  (2006) 096}, [\href{http://arxiv.org/abs/hep-th/0504110}{{\tt
  hep-th/0504110}}].

\bibitem{Hanany:2010zz}
A.~Hanany and A.~Zaffaroni, \emph{{The master space of supersymmetric gauge
  theories}}, \href{http://dx.doi.org/10.1155/2010/427891}{\emph{Adv.High
  Energy Phys.} {\bf 2010} (2010) 427891}.

\bibitem{Forcella:2008bb}
D.~Forcella, A.~Hanany, Y.-H. He and A.~Zaffaroni, \emph{{The Master Space of
  N=1 Gauge Theories}},
  \href{http://dx.doi.org/10.1088/1126-6708/2008/08/012}{\emph{JHEP} {\bf 0808}
  (2008) 012}, [\href{http://arxiv.org/abs/0801.1585}{{\tt 0801.1585}}].

\bibitem{Forcella:2008eh}
D.~Forcella, A.~Hanany, Y.-H. He and A.~Zaffaroni, \emph{{Mastering the Master
  Space}},
  \href{http://dx.doi.org/10.1007/s11005-008-0255-6}{\emph{Lett.Math.Phys.}
  {\bf 85} (2008) 163--171}, [\href{http://arxiv.org/abs/0801.3477}{{\tt
  0801.3477}}].

\bibitem{Forcella:2009bv}
D.~Forcella, \emph{{Master Space and Hilbert Series for N=1 Field Theories}},
  \href{http://arxiv.org/abs/0902.2109}{{\tt 0902.2109}}.

\bibitem{Zaffaroni:2008zz}
A.~Zaffaroni, \emph{{The master space of N=1 quiver gauge theories: Counting
  BPS operators}}, {\emph{8th Workshop on Continuous Advances in QCD
  (CAQCD-08)} (2008) 240--251}.

\bibitem{Forcella:2008ng}
D.~Forcella, A.~Hanany and A.~Zaffaroni, \emph{{Master Space, Hilbert Series
  and Seiberg Duality}},
  \href{http://dx.doi.org/10.1088/1126-6708/2009/07/018}{\emph{JHEP} {\bf 0907}
  (2009) 018}, [\href{http://arxiv.org/abs/0810.4519}{{\tt 0810.4519}}].

\bibitem{fulton}
W.~Fulton, \emph{{Introduction to toric varieties}}.
\newblock Annals of mathematics studies. Princeton Univ. Press, Princeton, NJ,
  1993.

\bibitem{cox1995homogeneous}
D.~A. Cox, \emph{The homogeneous coordinate ring of a toric variety},
  {\emph{arXiv preprint alg-geom/9210008} (1995) }.

\bibitem{sturmfels1996grobner}
B.~Sturmfels, \emph{Grobner bases and convex polytopes}, vol.~8.
\newblock American Mathematical Soc., 1996.

\bibitem{Benvenuti:2006qr}
S.~Benvenuti, B.~Feng, A.~Hanany and Y.-H. He, \emph{{Counting BPS operators in
  gauge theories: Quivers, syzygies and plethystics}},
  \href{http://dx.doi.org/10.1088/1126-6708/2007/11/050}{\emph{JHEP} {\bf 11}
  (2007) 050}, [\href{http://arxiv.org/abs/hep-th/0608050}{{\tt
  hep-th/0608050}}].

\bibitem{Feng:2007ur}
B.~Feng, A.~Hanany and Y.-H. He, \emph{{Counting Gauge Invariants: the
  Plethystic Program}},
  \href{http://dx.doi.org/10.1088/1126-6708/2007/03/090}{\emph{JHEP} {\bf 03}
  (2007) 090}, [\href{http://arxiv.org/abs/hep-th/0701063}{{\tt
  hep-th/0701063}}].

\bibitem{Butti:2006au}
A.~Butti, D.~Forcella and A.~Zaffaroni, \emph{{Counting BPS baryonic operators
  in CFTs with Sasaki- Einstein duals}},
  \href{http://dx.doi.org/10.1088/1126-6708/2007/06/069}{\emph{JHEP} {\bf 06}
  (2007) 069}, [\href{http://arxiv.org/abs/hep-th/0611229}{{\tt
  hep-th/0611229}}].

\bibitem{Butti:2007jv}
A.~Butti, D.~Forcella, A.~Hanany, D.~Vegh and A.~Zaffaroni, \emph{{Counting
  Chiral Operators in Quiver Gauge Theories}},
  \href{http://dx.doi.org/10.1088/1126-6708/2007/11/092}{\emph{JHEP} {\bf 0711}
  (2007) 092}, [\href{http://arxiv.org/abs/0705.2771}{{\tt 0705.2771}}].

\bibitem{hanany2007counting}
A.~Hanany, \emph{{Counting BPS operators in the chiral ring: The plethystic
  story}}, \href{http://dx.doi.org/10.1063/1.2803801}{\emph{AIP Conf. Proc.}
  {\bf 939} (2007) 165--175}.

\bibitem{Feng:2000mi}
B.~Feng, A.~Hanany and Y.-H. He, \emph{{D-brane gauge theories from toric
  singularities and toric duality}},
  \href{http://dx.doi.org/10.1016/S0550-3213(00)00699-4}{\emph{Nucl. Phys.}
  {\bf B595} (2001) 165--200}, [\href{http://arxiv.org/abs/hep-th/0003085}{{\tt
  hep-th/0003085}}].

\bibitem{Gulotta:2008ef}
D.~R. Gulotta, \emph{{Properly ordered dimers, $R$-charges, and an efficient
  inverse algorithm}},
  \href{http://dx.doi.org/10.1088/1126-6708/2008/10/014}{\emph{JHEP} {\bf 10}
  (2008) 014}, [\href{http://arxiv.org/abs/0807.3012}{{\tt 0807.3012}}].

\bibitem{Davey:2010px}
J.~Davey, A.~Hanany and R.-K. Seong, \emph{{Counting Orbifolds}},
  \href{http://dx.doi.org/10.1007/JHEP06(2010)010}{\emph{JHEP} {\bf 06} (2010)
  010}, [\href{http://arxiv.org/abs/1002.3609}{{\tt 1002.3609}}].

\bibitem{Hanany:2010ne}
A.~Hanany and R.-K. Seong, \emph{{Symmetries of Abelian Orbifolds}},
  \href{http://dx.doi.org/10.1007/JHEP01(2011)027}{\emph{JHEP} {\bf 01} (2011)
  027}, [\href{http://arxiv.org/abs/1009.3017}{{\tt 1009.3017}}].

\bibitem{DAuria:1983sda}
R.~D'Auria, P.~Fre and P.~van Nieuwenhuizen, \emph{{$N=2$ Matter Coupled
  Supergravity From Compactification on a Coset $G/H$ Possessing an Additional
  Killing Vector}},
  \href{http://dx.doi.org/10.1016/0370-2693(84)92018-5}{\emph{Phys. Lett. B}
  {\bf 136} (1984) 347--353}.

\bibitem{1986PhR...130....1D}
M.~J. {Duff}, B.~E.~W. {Nilsson} and C.~N. {Pope}, \emph{{Kaluza-Klein
  supergravity}},
  \href{http://dx.doi.org/10.1016/0370-1573(86)90163-8}{\emph{Physics Reports}
  {\bf 130} (Jan., 1986) 1--142}.

\bibitem{Nilsson:1984bj}
B.~Nilsson and C.~Pope, \emph{{Hopf Fibration of Eleven-dimensional
  Supergravity}},
  \href{http://dx.doi.org/10.1088/0264-9381/1/5/005}{\emph{Class.Quant.Grav.}
  {\bf 1} (1984) 499}.

\bibitem{Martelli:2008rt}
D.~Martelli and J.~Sparks, \emph{{Notes on toric Sasaki-Einstein
  seven-manifolds and AdS(4) / CFT(3)}},
  \href{http://dx.doi.org/10.1088/1126-6708/2008/11/016}{\emph{JHEP} {\bf 11}
  (2008) 016}, [\href{http://arxiv.org/abs/0808.0904}{{\tt 0808.0904}}].

\bibitem{Gauntlett:2004hh}
J.~P. Gauntlett, D.~Martelli, J.~F. Sparks and D.~Waldram, \emph{{A New
  infinite class of Sasaki-Einstein manifolds}},
  \href{http://dx.doi.org/10.4310/ATMP.2004.v8.n6.a3}{\emph{Adv. Theor. Math.
  Phys.} {\bf 8} (2004) 987--1000},
  [\href{http://arxiv.org/abs/hep-th/0403038}{{\tt hep-th/0403038}}].

\bibitem{Witten:1993yc}
E.~Witten, \emph{{Phases of N = 2 theories in two dimensions}},
  \href{http://dx.doi.org/10.1016/0550-3213(93)90033-L}{\emph{Nucl. Phys.} {\bf
  B403} (1993) 159--222}, [\href{http://arxiv.org/abs/hep-th/9301042}{{\tt
  hep-th/9301042}}].

\bibitem{M2}
D.~R. Grayson and M.~E. Stillman, ``Macaulay2, a software system for research
  in algebraic geometry.'' Available at
  \url{http://www.math.uiuc.edu/Macaulay2/}.

\bibitem{Gray:2008zs}
J.~Gray, Y.-H. He, A.~Ilderton and A.~Lukas, \emph{{STRINGVACUA: A Mathematica
  Package for Studying Vacuum Configurations in String Phenomenology}},
  \href{http://dx.doi.org/10.1016/j.cpc.2008.08.009}{\emph{Comput. Phys.
  Commun.} {\bf 180} (2009) 107--119},
  [\href{http://arxiv.org/abs/0801.1508}{{\tt 0801.1508}}].

\bibitem{Morrison:1998cs}
D.~R. Morrison and M.~R. Plesser, \emph{{Nonspherical horizons. 1.}},
  {\emph{Adv.Theor.Math.Phys.} {\bf 3} (1999) 1--81},
  [\href{http://arxiv.org/abs/hep-th/9810201}{{\tt hep-th/9810201}}].

\bibitem{Hanany:2014dia}
A.~Hanany and R.~Kalveks, \emph{{Highest Weight Generating Functions for
  Hilbert Series}},
  \href{http://dx.doi.org/10.1007/JHEP10(2014)152}{\emph{JHEP} {\bf 10} (2014)
  152}, [\href{http://arxiv.org/abs/1408.4690}{{\tt 1408.4690}}].

\bibitem{Candelas:1989ug}
P.~Candelas, P.~S. Green and T.~Hubsch, \emph{{Rolling Among Calabi-Yau
  Vacua}}, \href{http://dx.doi.org/10.1016/0550-3213(90)90302-T}{\emph{Nucl.
  Phys. B} {\bf 330} (1990) 49}.

\bibitem{Candelas:1989js}
P.~Candelas and X.~C. de~la Ossa, \emph{{Comments on Conifolds}},
  \href{http://dx.doi.org/10.1016/0550-3213(90)90577-Z}{\emph{Nucl. Phys. B}
  {\bf 342} (1990) 246--268}.

\bibitem{stanley1978hilbert}
R.~P. Stanley, \emph{Hilbert functions of graded algebras}, {\emph{Advances in
  Mathematics} {\bf 28} (1978) 57--83}.

\bibitem{Hanany:2012vc}
A.~Hanany and R.-K. Seong, \emph{{Brane Tilings and Specular Duality}},
  \href{http://dx.doi.org/10.1007/JHEP08(2012)107}{\emph{JHEP} {\bf 1208}
  (2012) 107}, [\href{http://arxiv.org/abs/1206.2386}{{\tt 1206.2386}}].

\end{thebibliography}\endgroup
%======================================================================

\end{document}